\documentclass[a4paper,12pt,nosumlimits]{amsart}
\usepackage{latexsym,amsmath,amssymb,enumerate,color,bbm,fullpage}
\usepackage[english]{babel}
\usepackage{amscd}

\title[The ADHM Construction of Noncommutative Instantons]{The ADHM Construction of Instantons \\[4pt] on Noncommutative Spaces}
\author{Simon Brain and Walter D. van Suijlekom}
\address{Institute for Mathematics, Astrophysics and Particle Physics, 
Faculty of Science, Radboud University Nijmegen,
Heyendaalseweg 135, 6525 AJ Nijmegen, The Netherlands}
\email{s.brain@math.ru.nl; waltervs@math.ru.nl}
 \numberwithin{equation}{section}

\date{26th August 2010}

\newtheorem{thm}{Theorem}[section]
\newtheorem{cor}[thm]{Corollary}
\newtheorem{lem}[thm]{Lemma}
\newtheorem{prop}[thm]{Proposition}
\newtheorem{rem}[thm]{Remark}
\theoremstyle{definition}
\newtheorem{defn}[thm]{Definition}
\newtheorem{example}[thm]{Example}

\newcommand{\half}{\tfrac{1}{2}}

\newcommand{\la}{\langle}
\newcommand{\ra}{\rangle}

\newcommand{\n}{\nabla}

\newcommand{\ep}{\epsilon}
\newcommand{\M}{\textup{M}}
\newcommand{\End}{\textup{End}}
\newcommand{\id}{\textup{id}}
\newcommand{\D}{\textup{d}}
\newcommand{\ii}{\mathrm{i}}
\newcommand{\p}{\partial}
\newcommand{\J}{\textrm{J}}
\newcommand{\bfa}{\mathbf{a}}
\newcommand{\bfb}{\mathbf{b}}
\newcommand{\bfc}{\mathbf{c}}
\newcommand{\bfd}{\mathbf{d}}
\newcommand{\Mod}{{\sf M}_k}

\newcommand{\qp}{{\sf q}}

\newcommand{\sfM}{\sf M}
\newcommand{\sfV}{{\sf V}}
\newcommand{\tsfV}{\widetilde\sfV}
\newcommand{\Pp}{{\sf P}}
\newcommand{\tPp}{\widetilde\Pp}
\newcommand{\Qp}{{\sf Q}}
\newcommand{\tQp}{\widetilde\Qp}
\newcommand{\trho}{\widetilde\rho}
\newcommand{\sa}{\mathrm{a}}
\renewcommand{\sb}{\mathrm{b}}
\renewcommand{\sc}{\mathrm{c}}
\newcommand{\sd}{\mathrm{d}}

\newcommand{\A}{\mathcal{A}}

\newcommand{\E}{\mathcal{E}}

\newcommand{\mH}{\mathcal{H}}
\newcommand{\mK}{\mathcal{K}}
\newcommand{\mL}{\mathcal{L}}

\newcommand{\Sp}{\textup{Sp}}
\newcommand{\GL}{\textup{GL}}
\newcommand{\U}{\textup{U}}
\newcommand{\SU}{\textup{SU}}

\newcommand{\HH}{\mathbb{H}}
\newcommand{\ZZ}{\mathbb{Z}}
\newcommand{\C}{\mathbb{C}}
\newcommand{\CP}{\mathbb{C}\mathbb{P}}
\newcommand{\TT}{\mathbb{T}}
\newcommand{\RR}{\mathbb{R}}
\newcommand{\R}{\mathbb{R}}

\newcommand{\mo}{{}^{\scriptscriptstyle(-1)}}
\newcommand{\bz}{{}^{{\scriptscriptstyle(0)}}}
\renewcommand{\o}{{}_{\scriptscriptstyle(1)}}
\renewcommand{\t}{{}_{\scriptscriptstyle(2)}}
\newcommand{\thr}{{}_{\scriptscriptstyle(3)}}
\newcommand{\fo}{{}_{\scriptscriptstyle(4)}}

 \def\lcross{{>\!\!\!\tl}}

\def\rbiprod{{\cdot\kern-.31em\tr\!\!\!<}}
\def\lbiprod{{>\!\!\!\triangleleft\kern-.33em\cdot}}

\newcommand{\utimes}{\,\underline{\otimes}\,}
\newcommand{\tr}{\triangleright}
\newcommand{\tl}{\triangleleft}

\begin{document}
\begin{abstract}
We present an account of the ADHM construction of
instantons on Euclidean space-time $\RR^4$ from the point of view of
noncommutative geometry. We recall the main ingredients of the
classical construction in a coordinate algebra format, which we then
deform using a cocycle twisting procedure to obtain a method for constructing families of instantons on noncommutative space-time, parameterised by solutions to an appropriate set of ADHM equations. We illustrate the
noncommutative construction in two special cases: the
Moyal-Groenewold plane $\RR^4_\hbar$ and the
Connes-Landi plane $\RR^4_\theta$.
\end{abstract}

\maketitle

\tableofcontents

\section{Introduction}\label{se:intro}
There has been a great deal of interest in recent years in the
construction of instanton gauge fields on space-times whose algebra
of coordinate functions is noncommutative. In classical geometry, an
instanton is a connection with anti-self-dual curvature on a smooth
 vector bundle over a four-dimensional manifold. The moduli space of
instantons on a classical four-manifold is an important invariant of
its differential structure \cite{dk} and it is only natural to try
to generalise this idea to the study of the differential geometry of
noncommutative four-manifolds.

In this article we study the construction of $\SU(2)$ instantons on
the Euclidean four-plane $\RR^4$ and its various noncommutative
generalisations. The problem of constructing instantons on classical
space-time was solved by the ADHM method of \cite{adhm:ci} and
consequently it is known that the moduli space of (framed) $\SU(2)$
instantons on $\RR^4$ with topological charge $k\in \ZZ$ is a
manifold of dimension $8k-3$. In what follows we review the ADHM
construction of instantons on classical $\RR^4$ and its extension to
noncommutative geometry. In particular, we study the construction of
instantons on two explicit examples of noncommutative Euclidean
space-time: the Moyal-Groenewold plane $\RR^4_\hbar$, whose algebra
of coordinate functions has commutation relations of the Heisenberg
form, and the Connes-Landi plane $\RR^4_\theta$ which arises as a
localisation of the noncommutative four-sphere $S^4_\theta$
constructed in \cite{cl:id} ({\em cf}. also \cite{cdv}). 

In order to deform the ADHM construction we adopt the technique of
`functorial cocycle twisting', a very general method which can be
used in particular to deform the coordinate algebra of any space
carrying an action of a locally compact Abelian group. Both of the
above examples of noncommutative space-times are obtained in this
way as deformations of classical Euclidean space-time: the Moyal
plane $\RR^4_\hbar$ as a twist along a group of translational
symmetries, the Connes-Landi plane $\RR^4_\theta$ as a twist along a
group of rotational symmetries. 

Crucially, the twisting technique does not just deform space-time alone: its functorial nature means that it simultaneously deforms any and all constructions which are covariant under the chosen group of symmetries. In particular, the parameter
spaces which occur in the ADHM construction also carry canonical actions of
the relevant symmetry groups, whence their coordinate algebras are also deformed
by the quantisation procedure. In this way, we obtain a method for constructing families of instantons which are parameterised by noncommutative
spaces.

As natural as this may seem, it immediately leads to the conceptual problem of how to interpret these spaces of noncommutative parameters. Indeed, our quest in each case is for a moduli space of instantons, which is necessarily modeled on the space of {\em all} connections on a given vector bundle over space-time \cite{ahs}. Even for noncommutative space-times, the set of such connections is an affine space and is therefore commutative \cite{ac:book}. To solve this problem, we adopt the strategy of \cite{bl:mod} and incorporate into the ADHM construction the `internal gauge symmetries' of noncommutative space-time \cite{ac:fncg}, in order to `gauge away' the noncommutativity and arrive at a classical space of parameters. 

The paper is organised as follows. The remainder of \S\ref{se:intro}
is dedicated to a brief review of the algebraic structures that we
shall need, including in particular the notions of Hopf algebras and their
comodules, together with the cocycle twisting construction itself. Following
this, in \S\ref{se:q-twi} and \S\ref{se:fams} we review the differential geometry of instantons from the point of view of coordinate algebras. We recall how to generalise these 
structures to incorporate the notions of noncommutative families of
instantons and their gauge theory.  In \S\ref{se:adhm} we sketch the ADHM construction of instanton from the point of view of coordinate algebras, stressing how the method is covariant under the
group of isometries of Euclidean space-time. It is precisely this covariance that we use in later sections to deform the ADHM construction by cocycle twisting. 

This coordinate-algebraic version of the ADHM construction has in fact already been studied in some detail \cite{bl:adhm,bl:mod}. In this sense, the first four sections of the present article consist mainly of review material. However, those earlier works studied the geometry of the ADHM construction from a somewhat abstract categorical point of view: in the present exposition, our goal is to give an explanation of the ADHM method in which we keep things as concrete as we are able. Moreover, the focus of these papers was on the construction of instantons on the Euclidean four-sphere $S^4$. In what follows we show how to adapt this technique to construct intantons on the local coordinate chart $\RR^4$. 

We illustrate the deformation procedure by giving the noncommutative ADHM
construction in two special cases. In \S\ref{se:moyal} we look at
how the method behaves under quantisation to give an ADHM
construction of instantons on the Moyal plane $\RR^4_\hbar$. On the other hand,
\S\ref{se:toric} addresses the issue of deforming the ADHM method to obtain a construction of instantons on the Connes-Landi plane $\RR^4_\theta$. It is these latter two sections which contain the majority of our new results.

Indeed, an algorithm for the construction of instantons on the Moyal plane $\RR^4_\hbar$ has been known for some time \cite{ns:ins}. However, a good understanding of the noncommutative-geometric origins of this construction has so far been lacking and our goal is to shed some light upon this subject (although it is worth pointing out that a different approach to the twistor theory of $\RR^4_\hbar$, from the point of view of noncommutative algebraic geometry, was carried out in \cite{kko}). Using the noncommutative twistor theory developed in \cite{bm:qtt}, we give an explicit construction of families of instantons on Moyal space-time,  from which we recover the well-known noncommutative ADHM equations of Nekrasov and Schwarz.

On the other hand, the noncommutative geometry of instantons and the ADHM construction on the Connes-Landi plane $\RR^4_\theta$ was investigated in \cite{lvs:pfns,lvs:nitcs,bl:adhm,bl:mod}. As already mentioned, however, this abstract geometric characterisation of the instanton construction is in need of a more concrete description. In the present paper we derive an explicit set of ADHM equations whose solutions parameterise instantons on the Connes-Landi plane (although we do not claim at this stage that {\em all} such instantons arise in this way).

We stress that our intention throughout the following is to present the various geometrical aspects of the ADHM construction in concrete terms. This means that, throughout the paper, we work purely at the algebraic level, {\em i.e.} with algebras of coordinate functions on all relevant spaces.  A more detailed approach, which in particular addresses all of the analytic aspects of the construction, will be presented elsewhere.

\subsection{Hopf algebraic structures}
 Let $H$ be a unital Hopf algebra. We denote its structure maps by
$$
\Delta:H\to H\otimes H,\qquad \ep:H\to \C, \qquad S:H\to H
$$
for the coproduct, counit and antipode, respectively. We use
Sweedler notation for the coproduct, $\Delta h=h\o \otimes h\t$, as
well as $(\Delta \otimes \id) \circ \Delta h = (\id \otimes \Delta)
\circ \Delta h = h\o \otimes h\t \otimes h \thr$ and so on, with
summation inferred.

We say that $H$ is {\em coquasitriangular} if it is equipped with a
convolution-invertible Hopf bicharacter $\mathcal{R}:H\otimes
H\rightarrow \C$ obeying
\begin{equation}\label{coq}
g\o h\o\mathcal{R}(h\t,g\t)=\mathcal{R}(h\o,g\o)h\t g\t
\end{equation}
for all $h,g\in H$. Convolution-invertibility means that there
exists a map $\mathcal{R}^{-1}:H\otimes H\rightarrow \C$ such that
\begin{equation}\label{con-inv}
\mathcal{R}(h\o,g\o)\mathcal{R}^{-1}(h\t,g\t)=\mathcal{R}^{-1}(h\o,g\o)\mathcal{R}(h\t,g\t)=\ep(g)\ep(h)
\end{equation}
for all $g,h \in H$. Being a Hopf bicharacter means that
\begin{equation}\label{hopf-bic}
\mathcal{R}(fg,h)=\mathcal{R}(f,h\o)\mathcal{R}(g,h\t), \qquad
\mathcal{R}(f,gh)=\mathcal{R}(f\o,h)\mathcal{R}(f\t,g),
\end{equation}
for all $f,g,h\in H$. If $\mathcal{R}$ also has the property that
$$
\mathcal{R}(h\o,g\o)\mathcal{R}(g\t,h\t)=\ep(g)\ep(h)
$$
for all $g,h\in H$, then we say that $H$ is {\em cotriangular}.

A vector space $V$ is said to be a {\em left $H$-comodule} if it is
equipped with a linear map $\Delta_V:V\to H\otimes V$ such that
$$
(\Delta_V\otimes
\id)\circ\Delta_V=(\Delta\otimes\id)\circ\Delta_V,\qquad
(\ep\otimes\id)\circ\Delta_V=\id.
$$
We shall often use the Sweedler notation $\Delta_V(v)=v\mo\otimes
v\bz$ for $v\in V$, again with summation inferred. If $V$, $W$ are
left $H$-comodules, a linear transformation $\sigma:V\to W$ is said
to be a {\em left $H$-comodule map} if it satisfies
$$
\Delta_W\circ\sigma=(\id\otimes \sigma)\circ\Delta_V.
$$
Given a pair $V,W$ of left $H$-comodules, the vector space $V\otimes
W$ is also a left $H$-comodule when equipped with the tensor product
$H$-coaction
\begin{equation}\label{mon}
\Delta_{V\otimes W}(v\otimes w)=v\mo w\mo \otimes \left(v\bz\otimes
w\bz\right)
\end{equation}
for each $v\in V$, $w \in W$. An algebra $A$ is said to be a {\em
left $H$-comodule algebra} if it is a left $H$-comodule equipped
with a product $m:A\otimes A\to A$ which is an $H$-comodule map,
meaning in this case that
$$
(\id\otimes m)\circ\Delta_{A\otimes A}=\Delta_A\circ m.
$$

Dually, a vector space $V$ is said to be a {\em left $H$-module} if
there is a linear map $\tr:H\otimes V\to V$, denoted $h\otimes
v\mapsto h\tr v$, such that
$$
h\tr(g\tr v)=(hg)\tr v,\qquad 1\tr v=v
$$
for all $v\in V$. An algebra $A$ is said to be a {\em left
$H$-module algebra} if it a left $H$-module equipped with a product
$m:A\otimes A$ which obeys
$$
h\tr(ab)=(h\o\tr a)(h\t \tr b)\qquad \text{for all} \quad a,b\in A.
$$
In the special case where $H$ is coquasitriangular, every left
$H$-comodule $V$ is also a left $H$-module when equipped with the
canonical left $H$-action
\begin{equation}\label{leftact}
\tr:H\otimes V\to V,\qquad h\tr v=\mathcal{R}(v\mo,h)v\bz
\end{equation}
for each $h\in H$, $v\in V$. The action \eqref{leftact} does not
commute with the $H$-coaction on $V$; rather it obeys the `crossed
module' condition
\begin{equation}\label{crossmod}
h\o v\mo\otimes h\t\tr v\bz=(h\o\tr v)\mo h\t\otimes (h\o \tr v)\bz
\end{equation}
for each $h \in H$ and $v \in V$. In particular, if $A$ is a left
$H$-comodule algebra then it is also a left $H$-module algebra when
equipped with the canonical action \eqref{leftact}.

To each left $H$-module algebra $A$ there is an associated {\em
smash product algebra} $A\lcross H$, which is nothing other than the
vector space $A\otimes H$ equipped with the product
\begin{equation}\label{cross}
(a\otimes h)(b\otimes g)=a(h\o\tr b)\otimes h\t g
\end{equation}
for each $a,b\in A$, $h,g\in H$. The main example of this
construction relevant to the present paper is the following.

\begin{example}\label{ex:smash}
Let $G$ be a locally compact Abelian Lie group with Pontryagin dual
group $\widehat G$ and let $\gamma:\widehat G\to \textrm{Aut}\,A$ be
an action of $\widehat G$  on a unital $*$-algebra $A$ by
$*$-automorphisms. Then associated to this action there is the crossed
product algebra $A\lcross_\gamma \,\widehat G$. On the other hand,
the $\widehat G$-action gives $A$ the structure of a $C_0(G)$-module
algebra and the Fourier transform on $\widehat G$ gives an
identification of $A\lcross_\gamma \,\widehat G$ with the smash
product $A\lcross \,C_0(G)$.
\end{example}

In this paper, our strategy is to keep things as explicit as
possible, avoiding the technical details of analytic arguments and
working purely at the algebraic level. For this reason, instead of
using the function algebra $C_0(G)$, we prefer to work with the
bialgebra $\A[G]$ of representative functions on $G$ equipped with
pointwise multiplication. In this setting, we can still make sense
of Example~\ref{ex:smash}: although the algebras $A\lcross_\gamma
\,\widehat G$ and $A\lcross \,C_0(G)$ are defined analytically using completions of appropriate function algebras, we think of the smash
product $A\lcross \,\A[G]$ as an `algebraic version' of the crossed
product algebra $A\lcross_\gamma \,\widehat G$. 

This also explains our use throughout the paper of coactions of Hopf algebras in place of group actions: the construction of crossed products by group actions is not defined at the algebraic level, whence we need to replace it by the smash product algebra instead. 

\subsection{Quantisation by cocycle twist}\label{se:nctt} Following \cite{ma:book}, in this
section we give a brief review of the deformation procedure that we
shall use later in the paper to `quantise' the ADHM construction of
instantons.

Let $H$ be a unital Hopf algebra whose antipode we assume to be
invertible. A {\em two-cocycle} on $H$ is a linear map $F:H\otimes
H\to \C$ which is unital, convolution-invertible in the sense of
Eq.~\eqref{con-inv}
%that there exists a linear map $F^{-1}:H\otimes H\to \C$ such that
%\begin{equation}\label{con-inv}
%F(h\o,g\o)F^{-1}(h\t,g\t)=F^{-1}(h\o,g\o)F(h\t,g\t)=\ep(g)\ep(h),\qquad
%h,g\in H,
%\end{equation}
and obeys the condition $\partial F=1$, {\em i.e.}
\begin{equation} \label{two-co}
F(g\o,f\o)F(h\o,g\o f\t)F^{-1}(h\t
g\thr,f\thr)F^{-1}(h\thr,g\fo)=\ep(f)\ep(h)\ep(g)
\end{equation}
for all $f,g,h \in H$. Given such an $F$, there is a twisted Hopf
algebra $H_F$ with the same coalgebra structure as $H$, but with
modified product $\bullet_{_F}$ and antipode $S_F$ given
respectively by
\begin{equation} \label{twprod}
h \bullet_{_F} g=F(h\o,g\o)h\t g\t F^{-1}(h\thr,g\thr),
\end{equation}
\begin{equation} \label{twanti}
S_F(h):=U(h\o)S(h\t)U^{-1}(h\thr),
\end{equation}
for each $h,g\in H_F$, where on the right-hand sides we use the
product and antipode of $H$ and define $U(h):=F(h\o,Sh\t)$. The
cocycle condition \eqref{two-co} is sufficient to ensure that the
product \eqref{twprod} is associative. In the case where $H$ is a
Hopf $*$-algebra %(so that in particular $\Delta$ is a $*$-algebra
%map and $(S\circ *)^2=\id$),
we also need to impose upon the cocycle $F$ the reality condition
\begin{equation} \label{real-co}
\overline{F(h,g)}=F((S^2g)^*,(S^2h)^*).
\end{equation}
In this situation the twisted Hopf algebra $H_F$ acquires a deformed
$*$-structure
\begin{equation} \label{tw-star}
h^{*_F}:=\overline{V^{-1}(S^{-1}h\o)}(h\t)^*\overline{V(S^{-1}h\thr)}
\end{equation}
for each $h\in H_F$, where $V(h):=U^{-1}(h\o)U(S^{-1}h\t)$.

If $H$ is a coquasitriangular Hopf algebra, then so is $H_F$. In
particular, if $H$ is commutative then $H_F$ is cotriangular with
`universal $R$-matrix' given by
\begin{equation}\label{R-braid}
\mathcal{R}(h,g):=F(g\o,h\o)F^{-1}(h\t,g\t)%\qquad\Psi_{V,W}(v\otimes w)=\mathcal{R}(w\mo,v\mo)w\bz\otimes v\bz
\end{equation}
for all $h,g\in H$. Since as coalgebras $H$ and $H_F$ are the same,
every left $H$-comodule is a left $H_F$-comodule and every
$H$-comodule map is an $H_F$-comodule map. This means that there is
an invertible functor which `functorially quantises' any
$H$-covariant construction to give an $H_F$-covariant one. As already mentioned, our
strategy will be to apply this idea to the construction of
instantons.

In passing from $H$ to $H_F$, from each left $H$-comodule algebra
$A$ we automatically obtain a left $H_F$-comodule algebra $A_F$
which as a vector space is the same as $A$ but has the modified
product
\begin{equation}\label{com-prod}
A_F\otimes A_F\to A_F, \qquad a\otimes b \mapsto a\cdot_F
b:=F(a\mo,b\mo)a\bz b\bz.
\end{equation}
If $A$ is a left $H$-comodule algebra and a $*$-algebra such that
the coaction is a $*$-algebra map, the twisted algebra $A_F$ also
has a new $*$-structure defined by
\begin{equation} \label{co-star}
a^{*_F}:=\overline{V^{-1}(S^{-1}a\mo)}(a\bz)^*
\end{equation}
for each $a \in A_F$.

\section{The Twistor Fibration}\label{se:q-twi}

The Penrose fibration $\CP^3\to S^4$ is an essential component of the ADHM
construction of instantons, since it encodes in its geometry the very
nature of the anti-self-duality equations on $S^4$ \cite{ma:gymf}. Following
\cite{bm:qtt,bl:adhm}, in this section we sketch the details of the Penrose
fibration from the point of view of coordinate algebras, then investigate what happens to the fibration upon passing to a local coordinate chart.

\subsection{The Penrose fibration}\label{se:hopf}
The $*$-algebra $\A[\C^4]$ of coordinate functions on the classical
space $\C^4$ is the commutative unital $*$-algebra generated by the
elements
$$\left\{z_j,z_l^*~|~j,l=1,\ldots,4\right\}.
$$
The coordinate algebra $\A[S^7]$ of the seven-sphere $S^7$ is the
quotient of $\A[\C^4]$ by the sphere relation
\begin{equation}\label{sev-sph}
z_1^*z_1+z_2^*z_2+z_3^*z_3+z_4^*z_4=1.
\end{equation}
%For later use, it is convenient to introduce the $\A[S^7]$-valued
%$4\times 2$ matrix
%\begin{equation}\label{u}
%\sfu:=\begin{pmatrix}z_1&z_2&z_3&z_4\\-%z_2^*&z_1^*&-%z^*_4&z_3^*\end{pmatrix}^{\textrm{tr}},
%\end{equation}
%with ${}^{\textrm{tr}}$ denoting matrix transposition. The sphere
%relation \eqref{sev-sph} is equivalent to imposing the condition
%$\sfu^*\sfu=\mathbbm{1}_2$, where $\mathbbm{1}_2$ denotes the
%$2\times 2$ identity matrix.
%
%In terms of the matrix \eqref{u}, there is a right action of the Lie
%group $\SU(2)$ on $\A[S^7]$ defined on generators by right matrix
%multiplication,
%\begin{equation}\label{su-act}
%\A[S^7]\times \SU(2)\to\A[S^7],\qquad \sfu \mapsto \sfu w, \qquad
%w=\begin{pmatrix}w_1&\bar w_2\\w_2&\bar w_1\end{pmatrix}\in\SU(2),
%\end{equation}
%extended to $\A[S^7]$ as a $*$-algebra map. The subalgebra
%$\textup{Inv}_{\SU(2)}\A[S^7]$ of invariant functions under the
%action \eqref{su-act} is generated as a unital $*$-algebra by the
%elements
On the other hand, the coordinate algebra $\A[S^4]$ of the four-sphere $S^4$ is the commutative unital $*$-algebra generated by the elements $x_1, x_1^*, x_2, x_2^*$ and $x_0=x_0^*$ subject to the sphere relation
$$
x_1^*x_1+x_2^*x_2+x_0^2=1.
$$
There is a canonical inclusion of algebras $\A[S^4]\hookrightarrow\A[S^7]$ defined on generators by
\begin{equation}\label{inc}
x_1=2(z_1z^*_3 + z_2^*z_4), \quad x_2=2(z_2z_3^* - z_1^*z_4),
\quad x_0=z_1z^*_1 + z_2z^*_2 - z_3z_3^* - z_4z_4^*,
\end{equation}
and extended as a $*$-algebra map.Clearly one has
\begin{equation}\label{four-sph}
x_1^*x_1+x_2^*x_2+x_0^2=(z_1^*z_1+z_2^*z_2+z_3^*z_3+z_4^*z_4)^2=1,
\end{equation}
so that the algebra inclusion is well-defined.
%
%whence the
%invariant subalgebra is precisely the coordinate algebra of a
%four-sphere,
%$$
%\textup{Inv}_{\SU(2)}\,\A[S^7]=\A[S^4].
%$$
%
%From the condition $\sfu^*\sfu=\mathbbm{1}_2$ it automatically
%follows that the matrix
%\begin{equation}\label{p}
%\pp:=\sfu\sfu^*=\tfrac{1}{2}\begin{pmatrix} 1+x_0 & 0 &x_1
%& -x_2^* \\ 0 & 1+x_0 & x_2 & x_1^* \\
%x_1^* & x_2^* & 1-x_0 & 0 \\ -x_2 & x_1 & 0 & 1-x_0\end{pmatrix}
%\end{equation}
%is a self-adjoint idempotent, {\em i.e.} $\pp^2=\pp=\pp^*$. Indeed,
%it is clear that the entries of $\pp$ generate an $\SU(2)$-invariant
%subalgebra, since we have
%$$
%(\sfu w)(\sfu w)^*=\sfu(ww^*)\sfu^*=\sfu\sfu^*.
%$$
%The equations \eqref{inc} define an inclusion of algebras
%$\A[S^4]\hookrightarrow\A[S^7]$, which is
This is just a coordinate algebra
description of the principal bundle $S^7\to S^4$ with structure
group $\SU(2)$ ({\em cf}. \cite{lvs:pfns} for further details of this construction).

\label{se:twistor} The twistor space of the Euclidean four-sphere
$S^4$ is nothing other than the complex projective space $\CP^3$. As
a real six-dimensional manifold, twistor space $\CP^3$ may be
identified with the set of all $4 \times 4$ Hermitian projector
matrices of rank one, since each such matrix uniquely determines and
is uniquely determined by a one-dimensional subspace of $\C^4$. Thus
the coordinate $*$-algebra $\A[\CP^3]$ of $\CP^3$ has a defining
matrix of (commuting) generators
\begin{equation}\label{q}
\qp:=\begin{pmatrix}a_1 & u_1 & u_2 & u_3\\ u_1^* & a_2 & v_3 & v_2 \\
u_2^* & v_3^* & a_3 & v_1 \\ u_3^* & v_2^* & v_1^* &
a_4\end{pmatrix} ,
\end{equation}
with $a_j^*=a_j$, $j=1,\ldots, 4$ and $\textup{Tr}\,\qp=\sum_j a_j
=1$, subject to the relations coming from the projection condition
$\qp^2=\qp$, that is to say $\sum_r \qp_{jr}\qp_{rl}=\qp_{jl}$ for
each $j,l=1,\ldots,4$.

%It is well-known that $\CP^3$ is obtained as a real manifold as the
%quotient of $S^7$ by the action of $\U(1)$. To see this in terms of
%coordinate algebras, we introduce the $\A[S^7]$-valued $4\times 1$
%matrix
%\begin{equation}\label{v}
%\sfv:=\begin{pmatrix}z_1&z_2&z_3&z_4\end{pmatrix}^{\textrm{t}}.
%\end{equation}
%In this case the sphere relation \eqref{sev-sph} is equivalent to
%the condition $\sfv^*\sfv=1$. Using this matrix, there is a right
%action of the Lie group $\U(1)$ on $\A[S^7]$ defined on generators
%for each $t\in \U(1)$ by
%$$
%\A[S^7]\times\U(1)\to \A[S^7],\qquad \sfv\mapsto \sfv t,
%$$
%extended as a $*$-algebra map. The subalgebra of invariant functions
%is isomorphic to $\A[\CP^3]$, being generated as a unital
%$*$-algebra {\em via} the identification of generators
There is a canonical inclusion of algebras $\A[\CP^3]\hookrightarrow\A[S^7]$ defined on generators by
\begin{equation}\label{tw-inc}
\qp_{jl}=z_jz_l^*,\qquad j,l=1,\ldots,4,
\end{equation}
with the relation $\qp_{11}+\qp_{22}+\qp_{33}+\qp_{44}=1$ coming
from the sphere relation \eqref{sev-sph}. 
%From the condition
%$\sfv^*\sfv=1$ we immediately recover that the matrix
%\begin{equation}\label{q2}
%\qp:=\sfv\sfv^*=(\qp_{jl})_{j,l=1,\ldots,4}
%\end{equation}
%is a self-adjoint idempotent, $\qp^2=\qp=\qp^*$, with trace equal to
%one. The identifications \eqref{tw-inc} define an algebra inclusion
%$\A[\CP^3]\hookrightarrow\A[S^7]$, which is just a coordinate
%algebra description of the principal bundle $S^7\to \CP^3$ with
%structure group $\U(1)$.

To determine the twistor fibration $\CP^3\to S^4$ in our coordinate
algebra framework, we need the map $\J:\A[\C^4]\to\A[\C^4]$ defined
on generators by
\begin{equation}\label{J}
\J(z_1,z_2,z_3,z_4):=(-z_2^*,z_1^*,-z_4^*,z_3^*)
\end{equation}
and extended as a $*$-anti-algebra map. %In doing so, we are
%identifying the quaternions $\HH$ with the set of $2\times 2$
%matrices
%$$
%c_1+c_2j\in \HH ~~\mapsto~~ \begin{pmatrix} c_1&-\bar c_2\\c_2&\bar
%c_1 \end{pmatrix}\in \M_2(\C).
%$$
%The map $\J$ corresponds at the level of coordinate algebras to
%right multiplication by the quaternion $j$. 
Equipping the algebra $\A[\C^4]$ with the map $\J$ thus identifies the space $\C^4$ with
the quaternionic vector space $\HH^2$ \cite{lprs:ncfi}. Accordingly, we define
$\A[\HH^2]$ to be the $*$-algebra $\A[\C^4]$ equipped with the
quaternionic structure $\J$.

Using the identification of generators \eqref{tw-inc}, the map $\J$
extends to an automorphism of the algebra $\A[\CP^3]$ given by
\begin{align*}
\J(a_1)&=a_2, & \J(a_2)&=a_1, & \J(a_3)&=a_4, & \J(a_4)&=a_3, &
\J(u_1)&=-u_1, \\ \J(v_1)&=-v_1, & \J(u_2)&=v_2^*, &
\J(u_3)&=-v_3^*, & \J(v_2)&=u_2^*, & \J(v_3)&=-u_3^*
\end{align*}
and extended as a $*$-anti-algebra map. It is straightforward to
check that the subalgebra of $\A[\CP^3]$ fixed by this automorphism
is precisely the four-sphere algebra $\A[S^4]$. Indeed, there is % that
%\begin{equation}\label{tw-ident}
%\pp=\sfu\sfu^*=\sfv\sfv^*+(\J\sfv)(\J\sfv)^*=\qp+\J(\qp)
%\end{equation}
%and hence 
an algebra inclusion $\A[S^4]\hookrightarrow\A[\CP^3]$
defined on generators by
\begin{equation}\label{tw-fib}
x_0\mapsto 2(a_1+a_2-1),\qquad x_1\mapsto 2(u_2+v_2^*),\qquad
x_2\mapsto 2(v_3-u_3^*),
\end{equation}
which is just a coordinate algebra description of the Penrose
fibration $\CP^3\to S^4$.

\subsection{Localisation of the twistor bundle}\label{se:local}
Next we look at what happens to the fibration
$\A[S^4]\hookrightarrow \A[\CP^3]$ when we pass to a local chart of
$S^4$ by removing a point. Indeed, it is well-known that in making
such a localisation the twistor bundle $\CP^3\to S^4$ becomes
isomorphic to the trivial fibration $\RR^4\times \CP^1$ over
$\RR^4$. In this section we illustrate this fact using the langauge
of coordinate algebras.

By definition, the localisation $\A_0[S^4]$ of $\A[S^4]$ is the
commutative unital $*$-algebra
$$
\A_0[S^4]:=\A[ \tilde x_1,\tilde x_1^*,\tilde x_2,\tilde
x_2^*,\tilde x_0,(1+\tilde x_0)^{-1}~|~\tilde x_1^*\tilde x_1+\tilde
x_2^*\tilde x_2+\tilde x_0^2=1,~(1+\tilde x_0)(1+\tilde
x_0)^{-1}=1].
$$
It is the algebra obtained from $\A[S^4]$ by adjoining an inverse
$(1+x_0)^{-1}$ to the function $1+x_0$; geometrically this
corresponds to `deleting' the point $(x_1,x_2,x_0)=(0,0,-1)$ from
the spectrum of the (smooth completion of the) algebra $\A[S^4]$,
with $\A_0[S^4]$ being the algebra of coordinate functions on the
resulting space. On the other hand, the coordinate algebra
$\A[\RR^4]$ of the Euclidean four-plane $\RR^4$ is the commutative
unital $*$-algebra generated by the elements
\begin{equation}\label{eucgen}
\left\{ \zeta_j,\zeta_l^*~|~j,l=1,2\right\}.
\end{equation}
Defining $|\zeta|^2:=\zeta^*_1\zeta_1+\zeta^*_2\zeta_2$, the element
$(1+|\zeta|^2)^{-1}$ clearly belongs to the (smooth completion of
the) algebra $\A[\RR^4]$ and so we have the following result.

\begin{lem}\label{st-proj}
The map $\A_0[S^4]\to \A[\RR^4]$ defined on generators by
\begin{equation}\label{chart}
\tilde x_1\mapsto 2\zeta_1(1+|\zeta|^2)^{-1},\qquad \tilde
x_2\mapsto 2\zeta_2(1+|\zeta|^2)^{-1},\qquad \tilde
x_0\mapsto(1-|\zeta|^2)(1+|\zeta|^2)^{-1}
\end{equation}
is a $*$-algebra isomorphism.\end{lem}

\proof The inverse of \eqref{chart} is the map
$\A[\RR^4]\to\A_0[S^4]$ given on generators by
\begin{equation}\label{ch-inv}
\zeta_1\mapsto \tilde x_1(1+\tilde x_0)^{-1},\qquad \zeta_2\mapsto
\tilde x_2(1+\tilde x_0)^{-1}
\end{equation}
and extended as a $*$-algebra map. Thus we have an isomorphism of
vector spaces. One checks that the elements $\tilde x_1$, $\tilde
x_2$, $\tilde x_0$ satisfy the same relation as the generators
$x_1$, $x_2$, $x_0$ of the algebra $\A[S^4]$. The difference is that
the point determined by the coordinate values
$(x_1,x_2,x_0)=(0,0,-1)$ is not in the spectrum of the (smooth)
algebra generated by the $\tilde x_1$, $\tilde x_2$, $\tilde x_0$.
In this way, we obtain $\RR^4$ as a local chart of the four-sphere
$S^4$, with the identification \eqref{chart} defining the `inverse
stereographic projection'. The point $(0,0,-1)$ will henceforth be
called the {\em point at infinity}.\endproof

At the level of twistor space $\CP^3$, passing to the local chart
$\RR^4$ by removing the point at infinity corresponds to removing
the fibre $\CP^1$ over that point: we refer to this copy of $\CP^1$
as the {\em line at infinity} and denote it by $\ell_\infty$. Under
the algebra inclusion \eqref{tw-fib}, inverting the element $1+x_0$
in $\A[S^4]$ is equivalent to inverting the element $a_1+a_2$ in
$\A[\CP^3]$. We denote by $\A_0[\CP^3]$ the resulting localised
algebra, {\em i.e.}
$$
\A_0[\CP^3]:=\A[
\qp_{jl},(a_1+a_2)^{-1}~|~\sum_r\qp_{jr}\qp_{rl}=\qp_{jl},~\textup{Tr}\,\qp=1,~(a_1+a_2)(a_1+a_2)^{-1}=1].
$$
We now show that this algebra is isomorphic to the algebra of
coordinate functions on the Cartesian product $\RR^4\times\CP^1$.

The coordinate algebra $\A[\CP^1]$ is the commutative unital
$*$-algebra generated by the entries of the matrix
\begin{equation}\label{CP-fib}
\tilde{\mathsf{q}}:=\begin{pmatrix}\tilde a_1&\tilde u_1\\\tilde
u_1^*&\tilde a_2\end{pmatrix}
\end{equation}
subject to the relations
$\tilde{\mathsf{q}}^2=\tilde{\mathsf{q}}^*=\tilde{\mathsf{q}}$ and
$\textup{Tr}\,\tilde{\mathsf{q}}=1$, that is to say $\tilde a_1\tilde
a_2=\tilde u_1^* \tilde u_1$, $\tilde a_1^*=\tilde a_1$, $\tilde
a_2^*=\tilde a_2$ and $\tilde a_1+\tilde a_2=1$. With the coordinate
algebra $\A[\RR^4]$ of \eqref{eucgen}, we have the following result.

\begin{lem}\label{le:triv}
There is a $*$-algebra isomorphism $\A[\RR^4]\otimes \A[\CP^1]\cong
\A_0[\CP^3]$ defined on generators by
$$
\zeta_1\otimes 1\mapsto (a_1+a_2)^{-1}(u_2+v_2^*),\qquad
\zeta_2\otimes 1\mapsto (a_1+a_2)^{-1}(v_3-u_3^*),
$$
$$
1\otimes \tilde a_1\mapsto (a_1+a_2)^{-1}a_1,\qquad 1\otimes \tilde
u_1\mapsto (a_1+a_2)^{-1}u_1,\qquad 1\otimes \tilde a_2\mapsto
(a_1+a_2)^{-1}a_2
$$
and extended as a $*$-algebra map.
\end{lem}

\proof We need to show that this map is an isomorphism of vector
spaces which respects the algebra relations in $\A[\RR^4]\otimes
\A[\CP^1]$. Using the expressions \eqref{inc} and \eqref{tw-inc}, we
find in $\A[S^7]$ the identities
$$
2(a_1+a_2)z_3=x_1^*z_1+x_2^*z_2,\qquad
2(a_1+a_2)z_3^*=x_1z_1^*+x_2z_2^*,
$$
$$
2(a_1+a_2)z_4=x_1z_2-x_2z_1,\qquad
2(a_1+a_2)z_4^*=x_1^*z_2^*-x_2^*z_1^*.
$$
In the localisation where $2(a_1+a_2)=1+x_0$ is invertible, these
expressions combined with the identifications $\qp_{ij}=z_iz_j^*$
define the inverse of the map stated in the lemma, so that we have a
vector space isomorphism. The algebra $\A[\CP^1]$ generated by
$\tilde a_1$, $\tilde a_2$, $\tilde u_1$ and $\tilde u_1^*$ is
identified with the subalgebra of $\A[\CP^3]$ generated by the
localised upper left $2\times 2$ block of the matrix \eqref{q}, {\em
i.e.} the subalgebra generated by the elements
$(a_1+a_2)^{-1}\qp_{ij}$ for $i,j=1,2$. It is easy to check that the
relations in $\A[\CP^1]$ are automatically preserved by this
identification. To check that the trace relation
$\textrm{Tr}\,\qp=\sum_j\qp_{jj}=1$ in $\A_0[\CP^3]$ also holds in
$\A[\RR^4]\otimes\A[\CP^1]$, one first computes that
$$
(a_1+a_2)^{-1}(z_1^*z_1+z_2^*z_2)\mapsto1\otimes 1,\qquad
(a_1+a_2)^{-1}(z_3^*z_3+z_4^*z_4)\mapsto
(\zeta_1^*\zeta_1+\zeta_2^*\zeta_2)\otimes 1,
$$
so that the trace relation holds if and only if
$(a_1+a_2)^{-1}\mapsto (1+|\zeta|^2)$, which is certainly true.
Moreover, in $\A[\CP^3]$ there are relations of the form
\begin{equation}\label{proj}
\qp_{ij}\qp_{kl}=z_iz_j^*z_kz_l^*=z_iz_l^*z_kz_j^*=\qp_{il}\qp_{kj}
\end{equation}
for $i,j,k,l=1,\ldots,4$. By adding together various linear
combinations and using the trace relation $\textrm{Tr}\,\qp=1$, one
finds that the relations \eqref{proj} are equivalent to the
projector relations $\qp^2=\qp$ . Hence it follows that the
projector relations in $\A_0[\CP^3]$ are equivalent to the remaining
relations in $\A[\RR^4]\otimes\A[\CP^1]$.\endproof

In this way, we see that there is a canonical inclusion of algebras
$\A[\RR^4]\hookrightarrow \A_0[\CP^3]$ in the obvious way; this is a
coordinate algebra description of the localised twistor fibration
$\RR^4\times\CP^1\to \RR^4$. Moreover, using the isomorphism in
Lemma~\ref{le:triv}, the quaternionic structure $\J$ of
Eq.~\eqref{J} is well-defined on the algebra $\A_0[\CP^3]$, with
$\A[\RR^4]$ being the $\J$-invariant subalgebra.

\subsection{Symmetries of the twistor fibration}\label{se:symms}
In later sections we shall obtain deformations of the twistor
fibration by the cocycle twisting of \S\ref{se:nctt}; for this we
need a group of symmetries acting upon the twistor bundle. Here we
describe the general strategy that we shall adopt.

We write $\M(2,\HH)$ for the algebra of $2\times 2$ matrices with
quaternion entries. The algebra $\A[\M(2,\HH)]$ of coordinate
functions on $\M(2,\HH)$ is the commutative unital $*$-algebra
generated by the entries of the $4\times 4$ matrix
\begin{equation} \label{defmat}A=\begin{pmatrix}{\sa}_{ij}&{\sb}_{ij}\\
{\sc}_{ij}&{\sd}_{ij}\end{pmatrix}=\begin{pmatrix}\alpha_1&-\alpha_2^*&\beta_1&-\beta_2^*\\\alpha_2&\alpha_1^*&\beta_2&\beta_1^*\\
\gamma_1&-\gamma_2^*&\delta_1&-\delta_2^*\\
\gamma_2&\gamma_1^*&\delta_2&\delta_1^*\end{pmatrix}.\end{equation}
We think of this matrix as being generated by a set of
quaternion-valued functions, writing
$$
\bfa=( {\sa}_{\textit{ij}} )=\begin{pmatrix}\alpha_1&-\alpha_2^*\\
\alpha_2&\alpha_1^*\end{pmatrix}$$ and similarly for the other
entries $\bfb,\bfc,\bfd$. The $*$-structure on this algebra is
evident from the matrix \eqref{defmat}. We equip $\A[\M(2,\HH)]$
with the matrix coalgebra structure
$$
\Delta(A_{ij})=\sum_r A_{ir}\otimes A_{rj}, \qquad
\ep(A_{ij})=\delta_{ij} \, \qquad \textup{for} \quad i,j=1,\ldots,4.
$$
Dual to the canonical action of $\M(2,\HH)$ on $\C^4\simeq\HH^2$
there is a left coaction defined by
\begin{equation}\label{conf-co}
\Delta_L:\A[\C^4]\to \A[\M(2,\HH)]\otimes\A[\C^4],\qquad z_j\mapsto
\sum_r A_{jr}\otimes z_r,
\end{equation}
extended as a $*$-algebra map. This coaction commutes with the
quaternionic structure \eqref{J}, in the sense that
$$
(\id\otimes \J)\circ\Delta_L=\Delta_L\circ \J,
$$
so that we have a coaction $\Delta_L:\A[\HH^2]\to
\A[\M(2,\HH)]\otimes\A[\HH^2]$ ({\em cf}. \cite{lprs:ncfi}).

The Hopf algebra $\A[\GL(2,\HH)]$ of coordinate functions on the
group $\GL(2,\HH)$ is obtained by adjoining to $\A[\M(2,\HH)]$ an
invertible group-like element $D$ obeying the relation
$D^{-1}=\textrm{det}\,A$, where $\textrm{det}\,A$ is the determinant
of the matrix \eqref{defmat}. This yields a left coaction
$$
\Delta_L:\A[\HH^2]\to \A[\GL(2,\HH)]\otimes\A[\HH^2],
$$
also defined by the formula \eqref{conf-co} and extended as a
$*$-algebra map.

The group $\GL(2,\HH)$ is the group of conformal symmetries of the
twistor fibration $\CP^3\to S^4$ \cite{pr:sst,mw:isdtt}. However, since we
are interested in the {\em localised} twistor bundle described in
Lemma~\ref{le:triv}, we work instead with the localised group of
symmetries $\GL^+(2,\HH)$, which is just the `coordinate patch' of
$\GL(2,\HH)$ in which the $2\times 2$ block $\bfa$ is assumed to be
invertible. The coordinate algebra $\A[\GL^+(2,\HH)]$ of this
localisation is obtained by adjoining to $\A[\GL(2,\HH)]$ an
invertible element $\widetilde D$ which obeys the relation
$\widetilde D^{-1}=\textrm{det}\,\bfa$.
%and $\textrm{det}(\sd-\sc \sa^{-1}\sb)$
The coaction of $\A[\GL^+(2,\HH)]$ on $\A[\HH^2]$ is once again
defined by the formula \eqref{conf-co}. We refer to \cite{bm:qtt}
for further details of this construction.

Throughout the paper, our strategy will be to deform the localised
twistor fibration and its associated geometry using the action of
certain subgroups of $\GL^+(2,\HH)$. In dual terms, we suppose $H$
to be a commutative Hopf $*$-algebra obtained {\em via} a Hopf algebra
projection
\begin{equation}\label{hopfsur}
\pi:\A[\GL^+(2,\HH)]\to H.
\end{equation}
This determines a left coaction of $H$ on $\A[\HH^2]$ by projection
of the coaction \eqref{conf-co}, namely
\begin{equation}\label{proj-co}
\Delta_\pi:\A[\HH^2]\to H\otimes \A[\HH^2],\qquad
\Delta_\pi:=(\pi\otimes\id)\circ\Delta_L,
\end{equation}
which makes $\A[\HH^2]$ into a left $H$-comodule $*$-algebra.
Moreover, we assume that this coaction respects the defining
relations of the localised twistor algebra $\A_0[\CP^3]$ given in
Lemma~\ref{le:triv}, whence it makes $\A_0[\CP^3]$ and $\A[\RR^4]$
into left $H$-comodule $*$-algebras in such a way that the algebra
inclusion $\A[\RR^4]\hookrightarrow \A_0[\CP^3]$ is a left
$H$-comodule map.

\section{Families of Instantons and Gauge Theory}\label{se:fams}
We are now ready to study differential structures on the twistor
fibration. In this section we recall the basic theory of
anti-self-dual connections on Euclidean space $\RR^4$ from the point
of view of noncommutative geometry. Following \cite{lprs:ncfi,bl:mod}, we then
generalise this by recalling what it means to have a {\em family} of
anti-self-dual connections on $\RR^4$ and when such families are
gauge equivalent. These notions will pave the way for the algebraic
formulation of the ADHM construction to follow.

\subsection{Differential structures and instantons}\label{se:diffcal}
As discussed, our intention is to present the construction of
connections and gauge fields in an entirely $H$-covariant framework,
from which all of our deformed versions will immediately follow by
functorial cocycle twisting. First of all we discuss the various
differential structures that we shall need.

We write $\Omega(\C^4)$ for the canonical differential calculus on
$\A[\C^4]$. It is the graded differential algebra generated by the
degree zero elements $z_j,z_l^*$, $j,l=1,\ldots, 4$, and the degree
one elements $\D z_j,\D z_l^*$, $j,l=1,\ldots,4$, subject to the
relations
$$
\D z_j\wedge \D z_l+\D z_l\wedge \D z_j=0,\qquad \D z_j\wedge \D
z_l^*+\D z_l^*\wedge \D z_j=0
$$
for $j,l=1,\ldots,4$. The exterior derivative $\D$ on $\Omega(\C^4)$
is defined by $\D:z_j\to \D z_j$ and extended uniquely using a
graded Leibniz rule. There is also an involution on $\Omega(\C^4)$
given by graded extension of the map $z_j\mapsto z_j^*$.

The story is similar for the canonical differential calculus
$\Omega(\RR^4)$. It is generated by the degree zero elements
$\zeta_j,\zeta_l^*$ and the degree one elements $\D\zeta_j$,
$\D\zeta_l^*$, $j,l=1,2$, subject to the relations
$$
\D \zeta_j\wedge \D \zeta_l+\D \zeta_l\wedge \D \zeta_j=0,\qquad \D
\zeta_j\wedge \D \zeta_l^*+\D \zeta_l^*\wedge \D \zeta_j=0.
$$
With $\pi:\A[\GL^+(2,\HH)]\to H$ a choice of Hopf algebra projection
as in Eq.~\eqref{hopfsur}, we assume throughout that the
differential calculi $\Omega(\C^4)$ and $\Omega(S^4)$ are (graded)
left $H$-comodule algebras such that the exterior derivative $\D$ is
a left $H$-comodule map, {\em i.e.} the coaction \eqref{proj-co}
obeys
$$
\Delta_\pi(\D z_j)=(\id\otimes\D)\Delta_\pi(z_j),\qquad
j=1,\ldots,4.
$$
In this way, the $H$-coactions on $\Omega(\C^4)$ and on
$\Omega(\RR^4)$ are given by extending the coaction on $\A[\C^4]$.

Next we come to discuss vector bundles over $\RR^4$. Of course, the
fact that $\RR^4$ is contractible means that the K-theory of the
algebra $\A[\RR^4]$ is trivial, {\em i.e.} all finitely generated
projective modules $\E$ over $\A[\RR^4]$ have the form
$\E=\A[\RR^4]^N$ for $N$ a positive integer and are equipped with a
canonical $\A[\RR^4]$-valued Hermitian structure
$\la\,\cdot\,|\,\cdot\,\ra$. A connection on $\E$ is a linear map
$\n:\E\to \E\otimes_{\A[\RR^4]} \Omega^1(\RR^4)$ satisfying the
Leibniz rule
$$
\n(\xi x)=(\n \xi)x + \xi \otimes \D x \qquad \text{for all} ~~\xi
\in \E, ~x \in \A[\RR^4].
$$
The connection $\n$ is said to be compatible with the Hermitian
structure on $\E$ if it obeys
$$
\la \n \xi|\eta\ra + \la\xi|\n \eta \ra= \D \la \xi|\eta \ra \qquad
\text{for all} ~~\xi, \eta \in \E, ~x \in \A[\RR^4].
$$
Since $\E$ is necessarily free as an $\A[\RR^4]$-module, any
compatible connection $\n$ can be written $\n=\D+\alpha$, where
$\alpha$ is a skew-adjoint element of
$\textup{Hom}_{\A[\RR^4]}(\E,\E\otimes_{\A[\RR^4]}
\Omega^1(\RR^4))$.

The curvature of $\n$ is the $\End_{\A[\RR^4]}(\E)$-valued two-form
$$
F:=\n^2=\D \alpha + \alpha^2.
$$
The Euclidean metric on $\RR^4$ determines the Hodge $*$-operator on
$\Omega(\RR^4)$, which on two-forms is a linear map
$*:\Omega^2(\RR^4)\to\Omega^2(\RR^4)$ such that $*^2=\id$. Since $H$
coacts by conformal transformations on $\A[\RR^4]$, there is an
$H$-covariant splitting of two-forms
$$
\Omega^2(\RR^4)=\Omega^2_+(\RR^4)\oplus \Omega^2_-(\RR^4)
$$
into self-dual and anti-self-dual components, {\em i.e.} the $\pm 1$
eigenspaces of the Hodge operator. The curvature $\n^2$ of a
connection $\n$ is said to be {\em anti-self-dual} if it satisfies
the equation $*F=-F$.

\begin{defn}
A compatible connection $\n$ on $\E$ is said to be an {\em
instanton} if its curvature $F=\n^2$ is an anti-self-dual two-form.
\end{defn}

The {\em gauge group} of $\E$ is defined to be
$$
\mathcal{U}(\E):=\left\{ U \in \textup{End}_{\A[\RR^4]}(\E)~|~ \la
U\xi|U\eta\ra=\la\xi|\eta\ra ~\text{for all}~ \xi,\eta \in \E
\right\}.
$$
It acts upon the space of compatible
connections by $$\n \mapsto \n^{U}:=U\n U^*$$ for each compatible
connection $\n$ and each element $U$ of $\mathcal{U}(\E)$. We say
that a pair of connections $\n_1$, $\n_2$ on $\E$ are {\em gauge
equivalent} if they are related by such a gauge transformation $U$.
The curvatures of gauge equivalent connections are related by
$F^U=(\n^U)^2=UFU^*$. Note in particular that if $\n$ has
anti-self-dual curvature then so does the gauge-transformed
connection $\n^U$.

We observe {\em a posteriori} that the above definitions
do not depend on the commutativity of the algebras $\A[\RR^4]$ and
$\Omega(\RR^4)$, so that they continue to make sense even if we allow
for deformations of the algebras $\A[\RR^4]$ and $\Omega(\RR^4)$.

\subsection{Noncommutative families of instantons}\label{se:ncfams}
Having given the definition of an instanton on $\RR^4$, we now come
to discuss what it means to have a {\em family} of instantons over
$\RR^4$. In the following, we let $A$ be an arbitrary (possibly
noncommutative) unital $*$-algebra.

\begin{defn}\label{de:fam-bun}
A {\em family of Hermitian vector bundles} over $\RR^4$
parameterised by the algebra $A$ is a finitely generated projective
right module $\E$ over the algebra $A\otimes\A[\RR^4]$ equipped with
an $A\otimes\A[\RR^4]$-valued Hermitian structure
$\la\,\cdot\,|\,\cdot\,\ra$.
\end{defn}

By definition, any such module $\E$ is given by a self-adjoint
idempotent $\Pp\in \textup{M}_N(A\otimes \A[\RR^4])$, {\em i.e.} an $N\times N$ matrix with entries in $A\otimes \A[\RR^4]$ satisfying
$\Pp^2=\Pp=\Pp^*$; the corresponding module is
$\E:=\Pp(A\otimes \A[\RR^4])^N$. Although
Definition~\ref{de:fam-bun} is given in terms of an arbitrary
algebra $A$, it is motivated by the case where $A$ is the
(commutative) coordinate algebra of some underlying classical space
$X$. In this situation, for each point $x\in X$ there is an
evaluation map $\textrm{ev}_x:A\to \C$ and the object
$$
\E_x:=(\textrm{ev}_x\otimes\id)\Pp((A\otimes \A[\RR^4])^N)
$$
is a finitely generated projective right $\A[\RR^4]$-module
corresponding to a vector bundle over $\RR^4$. In this way, the
projection $\Pp$ defines a family of Hermitian vector bundles
parameterised by the space $X$. When the algebra $A$ is
noncommutative, there need not be enough evaluation maps available,
but we may nevertheless work with the whole family at once.

Next we come to say what it means to have a family of connections
over $\RR^4$. We write $A\otimes\Omega^1(\RR^4)$ for the tensor
product bimodule over the algebra $A\otimes \A[\RR^4]$ and extend
the exterior derivative $\D$ on $\A[\RR^4]$ to $A\otimes \A[\RR^4]$
as $\id\otimes \D$.

\begin{defn}\label{de:fam-con}
A {\em family of connections} parameterised by the algebra $A$
consists of a family of Hermitian vector bundles
$\E:=\Pp(A\otimes\A[\RR^4])^N$ over $\A[\RR^4]$, together with a
linear map
$$
\n:\E\to \E\otimes_{A\otimes \A[\RR^4]} (A\otimes\Omega^1(\RR^4))
\simeq \E \otimes_{\A[\RR^4]} \Omega^1(\RR^4)
$$
obeying the Leibniz rule
$$
\n(\xi x)=(\n \xi)x + \xi \otimes (\id\otimes\D) x
$$
for all $\xi \in \E$ and $x \in A\otimes \A[\RR^4]$. The family is
said to be {\em compatible} with the Hermitian structure if it obeys
$\la \n \xi|\eta\ra + \la\xi|\n \eta \ra= (\id\otimes\D) \la
\xi|\eta \ra$ for all $\xi \in \E$ and $x \in A\otimes \A[\RR^4]$.
\end{defn}

It is clear that a family of connections parameterised by $A=\C$
({\em i.e.} by a one-point space) is just a connection in the usual
sense. In general, a given family of Hermitian vector bundles
$\E=\Pp(A\otimes \A[\RR^4])^N$ always carries the family of
Grassmann connections defined by
$$
\n_0=\Pp\circ (\textup{id}\otimes \D).
$$
It follows that any family of connections can be written in the form
$\n=\n_0+\alpha$, where $\alpha$ is a skew-adjoint element of
$$
\End_{A\otimes\A[\RR^4]} (\E,\E\otimes_{A\otimes \A[\RR^4]}
(A\otimes\Omega^1(\RR^4))) \simeq \End_{A\otimes\A[\RR^4]} (\E, \E
\otimes_{\A[\RR^4]} \Omega^1(\RR^4)).
$$

\begin{defn}\label{equiv fam}
Let $\E:=\Pp(A\otimes\A[\RR^4])^N$ be a family of Hermitian vector
bundles parameterised by the algebra $A$. The {\em gauge group} of
$\E$ is
$$\mathcal{U}(\E):=\{U\in \End_{A\otimes\A[\RR^4]}(\E)~|~\la
U\xi|U\eta\ra = \la\xi|\eta\ra\, ~\text{for all}~\xi,\eta\in\E\}.$$
We say that two families of compatible connections $\n_1$, $\n_2$ on
$\E$ are {\em equivalent families} and write $\n_1\sim \n_2$ if they
are related by the action of the unitary group, {\em i.e.} there
exists $U\in \mathcal{U}(\E)$ such that $\n_2=U\n_1U^*$.

More generally, if $\nabla_1$ and $\nabla_2$ are two families of connections on $\E$ parameterised by algebras $A_1$ and $A_2$ respectively, we say that $\nabla_1 \sim \nabla_2$ if there exists an algebra $B$ and algebra maps $\phi_1:A_1 \to B$ and $\phi_2:A_2 \to B$ such that $\phi_1^* \nabla_1 \sim \phi_2^* \nabla_2$ in the above sense. Here $\phi_i^* \nabla_i$ is the connection on $\E \otimes_{A_i}  B$ naturally induced by $\nabla_i$ ({\em cf}. \cite{bl:mod} for more details).
\end{defn}

In the case where $A=\C$, {\em i.e.} for a family parameterised by a
one-point space, the above relation reduces to the usual definition
of gauge equivalence of connections. In the case where the families
$\n_1$, $\n_2$ are Grassmann families associated to projections
$\Pp_1,\Pp_2\in\M_N(A\otimes\A[\RR^4])$, equivalence means that
$\Pp_2=U\Pp_1U^*$ for some unitary $U$.

\begin{lem}
With $\E=\Pp(A\otimes\A[\RR^4])^N$, there exists $\Pp_A\in
\M_N(A)$ such that there is an algebra isomorphism
$$
\End_{A\otimes\A[\RR^4]}(\E)\simeq \End_A(\Pp_A(A^N))
\otimes \A[\R^4]
$$
and hence an isomorphism $\mathcal{U}(\E)\simeq
\mathcal{U}(\End_A(\Pp_A(A^N)) \otimes \A[\R^4])$ of gauge
groups.
\end{lem}

\proof Since $\RR^4$ is topologically trivial, there is an
isomorphism of K-groups $\textup{K}_0(A\otimes\A[\RR^4])\cong
\textup{K}_0(A)$. It follows that for each projection
$\Pp\in\M_N(A\otimes\A[\RR^4])$ there exists a projection
$\Pp_A\in \M_N(A)$ such that $\Pp$ and
$\Pp_A\otimes 1$ are equivalent projections in
$\M_N(A\otimes\A[\RR^4])$. This implies that there is an isomorphism
\begin{equation}\label{triv}
\E=\Pp(A\otimes\A[\RR^4])^N\simeq \Pp_A(A^N) \otimes
\A[\R^4]
\end{equation}
of right $A\otimes \A[\RR^4]$-modules, from which the result follows
immediately.\endproof

With these ideas in mind, finally we arrive at the following definition of a family of instantons over $\RR^4$ parameterised by a (possibly noncommutative) $*$-algebra $A$.

\begin{defn}\label{fam inst}
A {\em family of instantons} over $\RR^4$ is a family of
compatible connections $\n$ over $\RR^4$ whose curvature $F:=\n^2$
obeys the anti-self-duality equation
$$
(\textup{id}\otimes *)F = -F,
$$
where $*$ is the Hodge operator on $\Omega^2(\RR^4)$.
\end{defn}

\section{The ADHM Construction}\label{se:adhm}
Next we review the ADHM construction of instantons on the classical
Euclidean four-plane $\RR^4$. We present the construction in a
coordinate algebra format which is covariant under the coaction of a
given Hopf algebra of symmetries, paving the way for a deformation
by the cocycle twisting of \S\ref{se:nctt}.

\subsection{The space of classical monads}\label{se:monad} We begin by describing the input data for the ADHM construction of
instantons. Although the ADHM construction is capable of
constructing instanton bundles of arbitrary rank, in this paper we
restrict our attention to the construction of vector bundles with
rank two.

\begin{defn}\label{de:monad}
Let $k\in\ZZ$ be a fixed positive integer. A {\em monad} over
$\A[\C^4]$ is a sequence of free right $\A[\C^4]$-modules,
\begin{equation}\label{monad}
0\to \mH\otimes \A[\C^4] \xrightarrow{\sigma_z} \mK\otimes \A[\C^4]
\xrightarrow{\tau_z} \mL\otimes \A[\C^4]\to 0,
\end{equation}
where $\mH$, $\mK$ and $\mL$ are complex vector spaces of dimensions
$k$, $2k+2$ and $k$ respectively, such that the maps $\sigma_z$,
$\tau_z$ are linear in the generators $z_1,\ldots,z_4$ of
$\A[\C^4]$. The first and last terms of the sequence are required to
be exact, so that the only non-trivial cohomology is in the middle
term.
\end{defn}

Given a monad \eqref{monad}, its cohomology
$\E:=\textrm{Ker}\,\tau_z/\textrm{Im}\,\sigma_z$ is a
finitely-generated projective right $\A[\C^4]$-module and hence
defines a vector bundle over $\C^4$. In fact, since the maps
$\sigma_z$, $\tau_z$ are assumed linear in the coordinate functions
$z_1,\ldots,z_4$, this vector bundle is well-defined over the
projective space $\CP^3$ \cite{oss:vb}.

With respect to ordered bases $(u_1,\ldots,u_k)$,
$(v_1,\ldots,v_{2k+2})$ and $(w_1,\ldots,w_k)$ for the vector spaces
$\mH$, $\mK$ and $\mL$ respectively, the maps $\sigma_z$ and
$\tau_z$ have the form
\begin{equation}\label{mod map}
\sigma_z: u_b\otimes Z \mapsto \sum\nolimits_{a,j} M_{ab}^j \otimes
v_a \otimes z_j Z,\qquad \tau_z:v_c\otimes Z\mapsto
\sum\nolimits_{d,j} N_{dc}^j \otimes w_d\otimes z_j,
\end{equation}
where $Z\in \A[\C^4]$ and the quantities $M^j:=(M^j_{ab})$ and
$N^j:=(N^j_{dc})$, $j=1,\ldots,4$, are complex matrices with
$a,c=1,\ldots,2k+2$ and $b,d=1,\ldots,k$. In more compact notation,
$\sigma_z$ and $\tau_z$ may be written
\begin{equation} \label{cpt-not}
\sigma_z=\sum\nolimits_j M^j \otimes z_j,\qquad
\tau_z=\sum\nolimits_j N^j \otimes z_j.
\end{equation}

It is immediate from the formul{\ae} \eqref{mod map} that the
composition $\tau_z\circ\sigma_z$ is given by
$$
\tau_z\circ\sigma_z:\mH\otimes \A[\C^4]\to \mL\otimes
\A[\C^4],\qquad u_b\otimes Z \mapsto \sum_{j,l,c,d}
N^j_{dc}M^l_{cb}\otimes w_d\otimes z_jz_lZ,
$$
with respect to the bases $(u_1,\ldots,u_k)$ and $(w_1,\ldots,w_k)$
of $\mH$ and $\mL$. It follows that the condition
$\tau_z\circ\sigma_z=0$ is equivalent to requiring that
\begin{equation}\label{adhm eqs}
\sum\nolimits_r \left(N^j_{dr}M^l_{rb}+N^l_{dr}M^j_{rb}\right)=0
\end{equation} for all $j,l=1,\ldots,4$ and
$b,d=1,\ldots,k$.

Introducing the conjugate matrix elements $M^j_{ab}{}^*$ and
$N^l_{dc}{}^*$, we use the compact notation
$(M^j{}^\dag)_{ab}=M^j_{ba}{}^*$ and
$(N^l{}^\dag)_{cd}=N^l_{dc}{}^*$. Then given a monad \eqref{monad},
the corresponding {\em conjugate monad} is defined to be
\begin{equation}\label{con-mon}
0\to \mL^*\otimes
\J(\A(\C^4_\theta))^\star\xrightarrow{\tau_{\J(z)}^*} \mK^*\otimes
\J(\A(\C^4_\theta))^*\xrightarrow{\sigma_{\J(z)}^*} \mH^*\otimes
\J(\A(\C^4_\theta))^*,
\end{equation}
where $\tau_{\J(z)}^*$ and $\sigma_{\J(z)}^*$ are the `adjoint' maps
defined by
$$\sigma_{\J(z)}^*=\sum_j M^j{}^\dag \otimes
\J(z_j)^*,\qquad \tau_{\J(z)}^*=\sum_j N^j{}^\dag \otimes \J(z_j)^*
$$
and $\J$ is the quaternionic involution defined in Eq.~\eqref{J} ({\em cf}. \cite{bl:adhm} for further explanation). If
a given monad \eqref{monad} is isomorphic to its conjugate
\eqref{con-mon} then we say that it is {\em self-conjugate}. A
necessary and sufficient condition for a monad to be self-conjugate
is that the maps $\sigma_z$ and $\tau_z$ should obey
$\tilde\tau_{J(z)}^*=-\tilde\sigma_z$ and
$\tilde\sigma_{J(z)}^*=\tilde\tau_z$, equivalently that the matrices
$M^j,$ $N^l$ should satisfy the reality conditions
\begin{equation}\label{monad star}
N^1=M^2{}^\dag, \quad N^2=-M^1{}^\dag, \quad N^3=M^4{}^\dag, \quad
N^4=-M^3{}^\dag.
\end{equation}

This is for fixed maps $\sigma_z$, $\tau_z$. In dual terms, by
allowing $\sigma_z$, $\tau_z$ to vary, we think of the elements
$M^j_{ab}$, $N^j_{dc}$ as coordinate functions on the space of all
possible pairs of $\A[\C^4]$-module maps
\begin{equation}\label{sig-tau}
\sigma_z:\mH\otimes \A[\C^4]\to \mK\otimes \A[\C^4],\qquad \tau_z:
\mK\otimes \A[\C^4] \to \mL \otimes \A[\C^4].
\end{equation}
Imposing the conditions \eqref{adhm eqs} and \eqref{monad star}, we
obtain coordinate functions on the space ${\sf M}_k$ of all
self-conjugate monads with index $k\in \ZZ$.

\begin{defn}\label{de:star-mon}
We write $\A[{\sf M}_k]$ for the commutative $*$-algebra generated
by the coordinate functions $M^j_{ab}$, $N^j_{dc}$ subject to the
relations \eqref{adhm eqs} and the $*$-structure \eqref{monad star}.
\end{defn}

\begin{rem}\label{re:framing}\textup{For each point $x\in {\sf M}_k$ there is an evaluation map
$$
\ep_x:\A[{\sf M}_k]\to \C
$$
and the complex matrices $(\ep_x\otimes\id)\sigma_z$ and
$(\ep_x\otimes\id)\tau_z$ define a self-conjugate monad over $\C^4$,
\begin{equation}
0\to \mH\otimes
\A[\C^4]\xrightarrow{(\textup{ev}_x\otimes\id)\sigma_z} \mK\otimes
\A[\C^4] \xrightarrow{(\textup{ev}_x\otimes\id)\tau_z} \mL\otimes
\A[\C^4]\to 0.
\end{equation}
As already remarked, the cohomology
$\E=\textrm{Ker},\tau_z/\textrm{Im}\,\sigma_z$ of a monad
\eqref{monad} defines a vector bundle $E$ over $\CP^3$; the
self-conjugacy condition \eqref{monad star} ensures that $E$ arises
{\em via} pull-back along the Penrose fibration $\CP^3\to S^4$. This
means that the bundle $E$ is trivial upon restriction to each of the
fibres of the Penrose fibration and, in particular, to the fibre
$\ell_\infty$ over the `point at infinity'.}
\end{rem}

As already mentioned, we wish to view our monads as being covariant
under a certain coaction of a Hopf algebra $H$. Recall that
$\A[\C^4]$ is already a left $H$-comodule algebra, with $H$-coaction
defined using the projection $\pi:\A[\GL^+(2,\HH)]\to H$ and the
formula \eqref{proj-co}. It automatically follows that the free
modules $\mH\otimes\A[\C^4]$, $\mK\otimes\A[\C^4]$ and $\mL\otimes
\A[\C^4]$ are also left $H$-comodules whose $\A[\C^4]$-module
structures are $H$-equivariant. It remains to address the
requirement that the module maps $\sigma_z$ and $\tau_z$ should be
$H$-equivariant as well.

\begin{lem}\label{le:mor}
The maps $\sigma_z:=\sum\nolimits_j M^j \otimes z_j$ and
$\tau_z:=\sum\nolimits_j N^j \otimes z_j$ are $H$-comodule maps if
and only if the coordinate functions $M^j_{ab}$, $N^l_{dc}$ carry
the left $H$-coaction
\begin{equation}\label{mon-co}
M^j_{ab}\mapsto \sum_r \pi (S(A_{rj}))\otimes M^r_{ab},\qquad
N^j_{dc}\mapsto \sum_r \pi (S(A_{rj}))\otimes N^r_{dc}
\end{equation}
for each $j=1,\ldots,4$ and $a,c=1,\ldots,2k+2$, $b,d=1,\ldots,k$.
\end{lem}

\proof Upon inspection of Eq.~\eqref{mod map} we see that $\sigma_z$
cannot possibly be an intertwiner for the $H$-coactions on
$\mH\otimes \A[\C^4]$ and $\mK\otimes \A[\C^4]$ unless we also allow
for a coaction of $H$ on the algebra $\A[{\sf M}_k]$ as well. Since
the definition of $\sigma_z$ depends only upon the generators
$M^j_{ab}$, it is enough to check equivariance only on these
generators. It is immediate that, for fixed $a,b$, the
four-dimensional $H$-comodule spanned by the generators $M^j_{ab}$,
$j=1,\ldots,4$, must be conjugate to the four-dimensional comodule
spanned by the generators $z_1,\ldots,z_4$, giving the coaction as
stated. Indeed, we verify that
\begin{align*}
\sum_{r}M^r\otimes z_r &\mapsto
\sum_{r,s}\pi(S(A_{sr})A_{rs})\otimes M^s\otimes
z_s\\
&=\sum_{s}\pi(\ep(A_{ss}))\otimes
M^s\otimes z_s\\
&=\sum_{s}1\otimes M^s\otimes z_{s},
\end{align*}
as required. The same analysis applies to the map $\tau_z$.\endproof

By extending it as a $*$-algebra map, the formula \eqref{mon-co}
equips $\A[{\sf M}_k]$ with the structure of a left $H$-comodule
$*$-algebra. This will be of paramount importance in later sections
when we come to deform the ADHM construction.

\subsection{The construction of instantons on $\RR^4$}\label{se:adhmcon} 
For self-conjugate monads, the maps of interest are the
$(2k+2)\times k$ algebra-valued matrices
\begin{align*}
\sigma_z &= M^1 \otimes z_1+M^2 \otimes z_2+M^3 \otimes z_3+M^4
\otimes z_4, \\
\sigma_{\J(z)} &=-M^1 \otimes z_2^* + M^2 \otimes z_1^* -M^3\otimes
z_4^* + M^4 \otimes z_3^*.
\end{align*}
In terms of these generators, the monad condition $\tau_z\sigma_z=0$
becomes $\sigma_{\J(z)}^* \sigma_z=0$. By polarisation of this
identity, one also finds that
$\sigma_{\J(z)}^*\sigma_{\J(z)}=\sigma_z^*\sigma_z$. The
identification of the vector space $\mK$ with its dual $\mK^*$ means
that the module $\mK \otimes \A[\C^4]$ acquires a bilinear form
given by
\begin{equation}\label{bil-form}
(\xi,\eta):=\la \J\xi|\eta\ra=\sum\nolimits_a (\J\xi)^*_a\eta_a
\end{equation}
for $\xi=(\xi_a)$ and $\eta=(\eta_a) \in \mK\otimes \A[\C^4]$, with
$\la \,\cdot\,|\,\cdot \,\ra$ the canonical Hermitian structure on
$\mK\otimes \A(\C^4_\theta)$. The monad condition $\sigma_{\J(z)}^*
\sigma_z=0$ implies that the columns of the matrix $\sigma_z$ are
orthogonal with respect to the form $(\,\cdot \,, \,\cdot \,)$.

Let us introduce the notation
\begin{equation}\label{rho}
\rho^2:=\sigma_z^*\sigma_z=\sigma_{\J(z)}^*\sigma_{\J(z)},
\end{equation}
a $k\times k$ matrix with entries in the algebra $\A[{\sf
M}_k]\otimes\A[\C^4]$. In order to proceed, we need this matrix
$\rho^2$ to be invertible, although of course this is not the case
in general. Thus we need to slightly enlarge the matrix
algebra $\M_k(\C)\otimes\A[{\sf M}_{k}]\otimes \A[\C^4]$ by
adjoining an inverse element $\rho^{-2}$ for $\rho^2$. Doing so is
equivalent to deleting a collection of points from the parameter
space $\Mod$, corresponding to the so-called `instantons of
zero-size' \cite{dk}. We henceforth assume that this has been done,
although we do not change our notation.

We collect together the matrices $\sigma_z$, $\sigma_{\J(z)}$ into
the $(2k+2)\times 2k$ matrix
\begin{equation}\label{matU}
\sfV :=\begin{pmatrix}\sigma_z & \sigma_{\J(z)}\end{pmatrix},
\end{equation}
which by the definition of $\rho^2$ obeys
$$
\sfV^* \sfV=\rho^2\begin{pmatrix} \mathbbm{1}_{k} & 0 \\ 0 &
\mathbbm{1}_k\end{pmatrix},
$$
where $\mathbbm{1}_k$ denotes the $k\times k$ identity matrix. We
form the matrix
\begin{equation}\label{proj-q}
\Qp:=\sfV\rho^{-2} \sfV^*=\sigma_z\rho^{-2}\sigma_z^* +
\sigma_{\J(z)}\rho^{-2}\sigma_{\J(z)}^*
\end{equation}
and for convenience we denote
\begin{equation}\label{q-decomp}
Q_z:=\sigma_z\rho^{-2}\sigma_z^*,\qquad
Q_{\J(z)}:=\sigma_{\J(z)}\rho^{-2}\sigma_{\J(z)}^*.
\end{equation}
Immediately we have the following result.

\begin{prop}\label{pr:adhm-pr}
The quantity $\Qp:=\sfV\rho^{-2} \sfV^*$ is a $(2k+2)\times(2k+2)$
projection, $\Qp^2=\Qp=\Qp^*$, with entries in the algebra
$\A[\Mod]\otimes \A[\RR^4]$ and trace equal to $2k$.
\end{prop}

\proof That $\Qp$ is a projection is a direct consequence of the
fact that $\sfV^*\sfV=\rho^2$. The matrices $Q_z$ and $Q_{\J(z)}$
are also projections: in fact they are orthogonal projections, since
$Q_zQ_{\J(z)}=0$. Moreover, both matrices $Q_z$ and $Q_{\J(z)}$ have
entries whose $\A[\C^4]$-components have the form $z_j^*z_l$ for
$j,l=1,\ldots,4$. From the proof of Lemma~\ref{le:triv}, we know
that we can rewrite each of these expressions in terms of generators
of the algebra $\A[\RR^4]\otimes\A[\CP^1]$. Since the matrix sum
$\Qp$ has entries which are $\J$-invariant, it follows that the
$\A[\C^4]$-components of these entries must lie in the
$\J$-invariant subalgebra of $\A[\RR^4]\otimes\A[\CP^1]$, which is
just $\A[\RR^4]$. For the trace, we compute that
\begin{align*}\textrm{Tr}
\,Q_z&=\sum\nolimits_\mu (\sigma_z\rho^{-2}
\sigma_z^*)_{\mu\mu}=\sum\nolimits_{\mu,r,s}(\sigma_z)_{\mu
r}(\rho^{-2})_{rs}(\sigma_z^*)_{s\mu}=\sum\nolimits_{\mu,r,s}(\rho^{-2})_{rs}(\sigma_z)_{\mu
r}(\sigma_z)^*_{s\mu}\\&=\sum\nolimits_{\mu,r,s}(\rho^{-2})_{rs}(\sigma_z)^*_{s\mu}(\sigma_z)_{\mu
r}=\sum\nolimits_{r,s}(\rho^{-2})_{rs}(\sigma_z^*\sigma_z)_{sr}=
\textup{Tr}\, \mathbbm{1}_k=k.
\end{align*}
A similar computation establishes that $Q_{\J(z)}$ also has trace
equal to $k$, whence the trace of $\Qp$ is $2k$ by
linearity.\endproof

From the projection $\Qp$ we construct the complementary projection
$\Pp:=\mathbbm{1}_{2k+2}-\Qp$, also having entries in the algebra
$\A[\Mod]\otimes \A[\RR^4]$. It is immediate that the trace of $\Pp$
is equal to two, so it follows that the finitely generated
projective right $\A[\Mod]\otimes\A[\RR^4]$-module
$$
\E:=\Pp(\A[\Mod]\otimes\A[\RR^4])^{2k+2}
$$
defines a family of rank two vector bundles over $\RR^4$
parameterised by the space $\Mod$ of self-conjugate monads.

We equip this family of vector bundles with the family of Grassmann
connections $\n:=\Pp\circ(\id\otimes \D)$. Immediately we obtain the
following result.

\begin{prop}\label{pr:asd}
The curvature $F=\Pp\left((\id\otimes \D)\Pp\right)^2$ of the family
of Grassmann connections $\n$ is anti-self-dual, that is
to say $(\id\otimes *)F=-F$.
\end{prop}

\proof By applying $\id\otimes \D$ to the relation
$\rho^{-2}\rho^2=\mathbbm{1}_{k}$ and using the Leibniz rule, one
finds that $(\id\otimes \D)\rho^{-2}=-\rho^{-2}((\id\otimes
\D)\rho^2)\rho^{-2}$ (this is a standard formula for calculating the
derivative of a matrix-valued function). Using this, one finds that
\begin{align*}
(\id\otimes \D)(\sfV \rho^{-2}\sfV^*)=\Pp((\id\otimes
\D)\sfV)\rho^{-2}\sfV^*+\sfV\rho^{-2}((\id\otimes \D)\sfV^*)\Pp,
\end{align*}
and hence in turn that
\begin{align*}
((\id\otimes \D)\Pp)\wedge((\id\otimes \D)\Pp)=\Pp((\id\otimes
\D)\sfV)\rho^{-2}&((\id\otimes \D)\sfV^*)\Pp\\
&+\sfV\rho^{-2}((\id\otimes \D)\sfV^*)\Pp((\id\otimes
\D)\sfV)\rho^{-2}\sfV^*,
\end{align*}
where we have used the facts that
$\rho^{-2}\sfV^*\Pp=0=\Pp\sfV\rho^{-2}$. The second term in the
above expression is identically zero when acting on any element in
the image $\E$ of $\Pp$, whence the curvature $F$ of the family $\n$
works out to be
\begin{align*}
F&=\Pp\left((\id\otimes \D)\Pp\right)^2\\&=\Pp((\id\otimes
\D)\sfV)\rho^{-2}((\id\otimes \D)\sfV^*)\Pp\\
&=\Pp\left(((\id\otimes \D)\sigma_z)\rho^{-2}((\id\otimes
\D)\sigma_z^*)+((\id\otimes \D)\sigma_{\J(z)})\rho^{-2}((\id\otimes
\D)\sigma_{\J(z)}^*)\right)\Pp.
\end{align*}
It is clear by inspection that on twistor space $\A_0[\CP^3]$ this $F$
is a horizontal two-form of type $(1,1)$ and it is known \cite{ma:gymf} that
such a two-form is necessarily the pull-back of an anti-self-dual
two-form on $\RR^4$.\endproof

Thus we have reproduced the ADHM construction of instantons on
$\RR^4$ in our coordinate algebra framework: as usual, we must now
address the question of the extent to which the construction depends
on the choice of bases for the vector spaces $\mH$, $\mK$, $\mL$
that we made in \S\ref{se:monad}. 

It is clear that we are free to
act on the $\A[\C^4]$-module $\mK\otimes \A[\C^4]$ by a unitary
element of the matrix algebra $\M_{2k+2}(\C)\otimes \A[\C^4]$. In
order to preserve the instanton construction, we must do so in a way
which preserves the bilinear form $(\cdot\,,\cdot)$ of
Eq.~\eqref{bil-form} determined by the identification of $\mK$ with
its dual $\mK^*$. It follows that the map $\sigma_z$ in
Eq.~\eqref{mod map} is defined up to a unitary transformation $U \in
\textup{End}_{\A[\C^4]}(\mathcal{K}\otimes \A[\C^4])$ which commutes
with the quaternion structure $\J$, namely the elements of the group
$$
\textup{Sp}(\mathcal{K}\otimes\A[\C^4]):=\left\{ U \in
\textup{End}_{\A(\C^4_\theta)}(\mathcal{K}\otimes\A[\C^4])~|~\la
U\xi|U\xi\ra=\la \xi|\xi\ra,~\J(U\xi)=U\J(\xi)\right\}.
$$
Similarly, we are free to change basis in the module
$\mathcal{H}\otimes \A[\C^4]$, whence the map $\tau_z$ in
Eq.~\eqref{mod map} is defined up to an invertible transformation
$W\in\GL(\mathcal{H}\otimes\A[\C^4])$. Given $U \in
\textup{Sp}(\mathcal{K}\otimes\A[\C^4])$ and $W \in
\GL(\mathcal{H}\otimes\A[\C^4])$, the available freedom in the ADHM
construction is to map $\sigma_z \mapsto U\sigma_zW$.

\begin{prop}\label{pr:gauge}
For all $W\in\GL(\mathcal{H})$ the projection
$\Pp=\mathbb{I}_{2k+2}-\Qp$ is invariant under the transformation
$\sigma_z\mapsto\sigma_z W$. For all $U\in
\textup{Sp}(\mathcal{K})$, under the transformation $\sigma_z
\mapsto U\sigma_z$ the projection $\Pp$ of transforms as $\Pp\mapsto
U\Pp U^*$.
\end{prop}

\proof One first checks that $\rho^2\mapsto (\sigma_zW)^*
(\sigma_zW) = W^* \rho^2 W$, so that
$$\Qp_z \mapsto \sigma_z W (W^* \rho^2 W)^{-1}W^*
\sigma_z^* = \sigma_z W (W^{-1} \rho^{-2} (W^*)^{-1}) W^* \sigma_z^*
= \Qp_z,$$ whence the projection $\Pp$ is unchanged. Replacing
$\sigma_z$ by $U \sigma_z$ leaves $\rho^2$ invariant (since $U$ is
unitary) and so has the effect that
$$\Qp_z \mapsto U\sigma_z \rho^{-2} \sigma_z^* U^* =
U\Qp_zU^*,$$ whence it follows that $\Pp$ is mapped to $U\Pp
U^*$.\endproof

In this way, such changes of module bases result in gauge equivalent families of instantons. However, from the point of view
of constructing equivalence classes of connections it is in fact sufficient
to consider the effect of the subgroups of `constant' module
automorphisms, {\em i.e.} those generated changes of basis in the
vector spaces $\mH$, $\mK$ and $\mL$,
described by the group $\textrm{Sp}(\mK)=\textrm{Sp}(k+1)\subset
\textrm{Sp}(\mK\otimes\A[\C^4])$ and the group
$\textrm{GL}(k,\RR)\subset\textrm{GL}(\mH\otimes \A[\C^4])$ \cite{adhm:ci}.

Although it is beyond our scope to prove this here, we note that the algebra
$\A[{\sf M}_k]$ has a total of $4k(2k+2)$ generators and $5k(k-1)$
constraints (determined by the orthogonality relations
$\sigma_{J(z)}^*\sigma_z=0$); the $\textrm{Sp}(k+1)$ symmetries
impose a further $(k+1)(2(k+1)+1)$ constraints and the
$\textrm{GL}(k,\RR)$ a further $k^2$ constraints. This elementary
argument yields that the construction has
$$(8k^2+8k)-5k(k-1)-(3k^2+5k+3)=8k-3$$
degrees of freedom, in precise agreement with the dimension of the
moduli space computed in \cite{ahs}.

\begin{defn}\label{de:mon-eq}
We say that a pair of self-conjugate monads are {\em equivalent} if they are related by a change of bases of the vector spaces $\mH$, $\mK$, $\mL$ of the above form, {\em i.e.}  by a pair of linear transformations $U\in \Sp(k+1)$ and $W\in\GL(k,\RR)$. We denote by $\sim$ the resulting equivalence relation on the space ${\sfM}_k$ of self-conjugate monads.
\end{defn}

\section{The Moyal-Groenewold Noncommutative Plane $\RR^4_\hbar$}
\label{se:moyal} The Moyal noncommutative space-time $\RR^4_\hbar$
is arguably one of the best-known and most widely-studied examples of a
noncommutative space. In this section we analyse the construction of
instantons on this space from the point of view of cocycle twisting.
In this section we show how to deform Euclidean space-time $\RR^4$ and its
associated geometric structure into that of the Moyal-Groenewold space-time;
then we look at what happens to the ADHM construction of instantons
under the deformation procedure.

\subsection{A Moyal-deformed family of monads}\label{se:moymon}
In order to deform the ADHM construction of instantons, we need to
choose a Hopf algebra $H$ of symmetries together with a two-cocycle
$F$ by which to perform the twisting. For our twisting Hopf algebra
we take $H=\A[\RR^4]$, the algebra of coordinate functions on the
additive group $\RR^4$. It is the commutative unital $*$-algebra
\begin{equation}\label{defhopf}
\A[\RR^4]=\A[t_j,t_j^*~|~j=1,2]
\end{equation}
equipped with the Hopf algebra structure
\begin{equation}\label{hopf-trans2}
\Delta(t_j)=1\otimes t_j+t_j\otimes 1,\qquad \ep(t_j)=0,\qquad
S(t_j)=-t_j,
\end{equation}
with $\Delta$, $\ep$ extended as $*$-algebra maps and $S$ extended
as a $*$-anti-algebra map. In order to deform the twistor fibration,
we have to equip our various algebras with left $H$-comodule algebra
structures, which we achieve using the discussion of
\S\ref{se:symms}. There is a Hopf algebra projection from
$\A[\GL^+(2,\HH)]$ onto $H$, defined on generators by
\begin{equation}\label{moy-proj}
\pi:\A[\GL^+(2,\HH)]\to H,\qquad  \begin{pmatrix}\alpha_1&-\alpha_2^*&\beta_1&-\beta_2^*\\\alpha_2&\alpha_1^*&\beta_2&\beta_1^*\\
\gamma_1&-\gamma_2^*&\delta_1&-\delta_2^*\\
\gamma_2&\gamma_1^*&\delta_2&\delta_1^*\end{pmatrix}\mapsto \begin{pmatrix}1&0&0&0\\0&1&0&0\\
t_1^*&t_2^*&1&0\\-t_2&t_1&0&1\end{pmatrix}
\end{equation}
and extended as a $*$-algebra map. Using Eq.~\eqref{proj-co}, this
projection determines a left $H$-coaction $\Delta_\pi:\A[\C^4]\to
H\otimes \A[\C^4]$ by
\begin{align}\label{tr-co1}
z_1&\mapsto 1\otimes z_1,\qquad z_3\mapsto t_1^*\otimes
z_1+t_2^*\otimes z_2+1\otimes z_3,\\ \label{tr-co2} z_2&\mapsto
1\otimes z_2,\qquad z_4\mapsto -t_2\otimes z_1+t_1\otimes
z_2+1\otimes z_4,
\end{align} extended as a $*$-algebra map. Using the
identification of generators \eqref{ch-inv}, the coordinate algebra
$\A[\RR^4]$ of Euclidean space therefore carries the coaction
\begin{equation}\label{tr-sph}
\A[\RR^4]\to H\otimes\A[\RR^4],\qquad \zeta_1\mapsto
1\otimes\zeta_1+t_1\otimes 1,\quad \zeta_2\mapsto 1\otimes
\zeta_2+t_2\otimes 1,
\end{equation}
making $\A[\RR^4]$ into a left $H$-comodule $*$-algebra.

Let $(\p_j{}^l)$, $j,l=1,2$, be the Lie algebra of translation
generators dual to $H$. Writing
$$
\tau:=(\tau_r{}^s)=\begin{pmatrix}t_1^*&t_2^*\\-t_2&t_1\end{pmatrix},\qquad
r,s=1,2,
$$
this means that there is a non-degenerate pairing
$$
\la \p_j{}^l,\tau_r{}^s\ra=\delta_j^s\,\delta_r^l,\qquad
j,l,r,s=1,2,
$$
which extends to an action on products of the generators $t_j,t_l^*$
by differentiation and evaluation at zero. Using this pairing, we
define a twisting two-cocycle by
\begin{equation*}
F:H\otimes H\to \C,\qquad F(h,g)=\left\la
\textrm{exp}\left(\tfrac{1}{2}\ii\Theta_r{}^{r'}{}_s{}^{s'}
\p_{r'}{}^r\otimes \p_{s'}{}^s\right),h\otimes g\right\ra
\end{equation*}
for $h,g\in H$, where $\Theta=(\Theta_r{}^{r'}{}_s{}^{s'})$,
$r,r',s,s'=1,2$, is a real $4\times 4$ anti-symmetric matrix with
rows $rr'$ and columns $ss'$, which we may choose to have the
canonical form
$$
\Theta=\hbar\begin{pmatrix}0&0&0&\alpha\\0&0&\beta&0\\0&-\beta&0&0\\-\alpha&0&0&0\end{pmatrix}
$$
for non-zero real constants $\alpha,\beta$ and $\hbar>0$ a
deformation parameter. We assume for simplicity that $\alpha+\beta\ne 0$. This $F$ is multiplicative ({\em i.e.} it is
a Hopf bicharacter in the sense of Eq.~\eqref{hopf-bic}) and so it
is determined by its values on the generators $(\tau_r{}^s)$,
$r,s=1,2$. One computes in particular that
$$
F(t_1^*,t_1)=\tfrac{1}{2}\alpha\ii\hbar,\qquad
F(t_2^*,t_2)=-\tfrac{1}{2}\beta\ii\hbar,
$$
with $F$ evaluating as zero on all other pairs of generators. From
the formul{\ae} \eqref{twprod}--\eqref{twanti} and \eqref{tw-star},
one immediately finds that $H=H_F$ as a Hopf $*$-algebra. However,
the effect of the twisting on the $H$-comodule algebras $\A[\C^4]$
and $\A[\RR^4]$ is not trivial, as shown by the following lemmata.

\begin{lem}\label{le:stdef}
The algebra relations in the $H$-comodule algebra $\A[\C^4]$ are
twisted into
\begin{align}\label{def-h1}
[z_3,z_4]&=\ii\hbar (\alpha+\beta)z_1z_2, &
[z_3^*,z_4^*]&=\ii\hbar(\alpha+\beta)z_1^*z_2^*,\\
\label{def-h2}[z_3,z_3^*]&=\ii\hbar \alpha z_1z_1^*-\ii\hbar \beta
z_2z_2^*,& [z_4,z_4^*]&=\ii\hbar \beta z_1z_1^*-\ii\hbar \alpha
z_2z_2^*,
\end{align}
with all other relations left unchanged. In particular, the
generators $z_1,z_2$ and their conjugates remain central in the
deformed algebra.
\end{lem}

\proof The cocycle-twisted product on the $H$-comodule algebra
$\A[\C^4]$ is defined by the formula \eqref{com-prod}; the
corresponding algebra relations can be expressed using the
`universal $R$-matrix' of Eq.~\eqref{R-braid}, namely
\begin{equation}\label{R-rels}
z_j\cdot_F z_l=\mathcal{R}(z_l\mo,z_j\mo)z_l\bz \cdot_F
z_j\bz,\qquad z_j\cdot_F
z_l^*=\mathcal{R}(z_l^*{}\mo,z_j\mo)z_l^*{}\bz\cdot_F z_j\bz;
\end{equation}
One finds in particular that the $R$-matrix has the values
\begin{equation}\label{eval}
\mathcal{R}(t_1^*,t_1)=2F^{-1}(t_1^*,t_1)=-\ii\hbar \alpha,\qquad
\mathcal{R}(t_2^*,t_2)=2F^{-1}(t_2^*,t_2)=\ii\hbar\beta,
\end{equation}
and gives zero when evaluated on all other pairs of generators. By
explicitly computing Eqs.~\eqref{R-rels} (and omitting the product
symbol $\cdot_F$), one finds the relations as stated in the lemma.
We denote by $\A[\C^4_\hbar]$ the $*$-algebra generated by
$\{z_j,z_j^*~|~j=1,\ldots,4\}$ modulo the relations
\eqref{def-h1}--\eqref{def-h2}. This makes $\A[\C^4_\hbar]$ into a
left $H_F$-comodule $*$-algebra.\endproof

\begin{lem}
The algebra relations in the $H$-comodule algebra $\A[\RR^4]$ are
twisted into
\begin{equation}\label{st-rels}
[\zeta_1^*,\zeta_1]=\ii\hbar \alpha ,\qquad
[\zeta_2^*,\zeta_2]=-\ii\hbar \beta,\qquad j,l=1,2,
\end{equation}
with vanishing commutators between all other pairs of generators.
\end{lem}

\proof The product in $\A[\RR^4]$ is twisted using the formula
\eqref{com-prod}. Once again omitting the product symbol $\cdot_F$,
the corresponding algebra relations are computed to be those as
stated. We denote by $\A[\RR^4_\hbar]$ the algebra generated by
$\zeta_1$, $\zeta_2$, $\zeta_1^*$, $\zeta_2^*$, modulo the relations
\eqref{st-rels}. This makes $\A[\RR^4_\hbar]$ into a left
$H_F$-comodule $*$-algebra.\endproof

\begin{rem}\label{re:mo-triv}\textup{Since the generators $z_1$, $z_2$ and their conjugates $z_1^*$, $z_2^*$ remain central in the algebra $\A[\C^4_\hbar]$, we see immediately from
Lemma~\ref{le:triv} that the localised twistor algebra has the form
$\A[\RR^4_\hbar]\otimes \A[\CP^1]$. Only the base $\RR^4_\hbar$ of
the twistor fibration is deformed; the typical fibre $\CP^1$ remains
classical.}\end{rem}

The canonical differential calculi described in \S\ref{se:diffcal}
are also deformed using this cocycle twisting procedure. The
relations in the deformed calculi are given in the following
lemmata.

\begin{lem}
The twisted differential calculus $\Omega(\C^4_\hbar)$ is generated
by the degree zero elements $z_j, z_l^*$ and the degree one elements
$\D z_j, \D z_l^*$ for $j,l=1,\ldots,4$, subject to the bimodule
relations between functions and one-forms
\begin{align*}
[z_3,\D z_4]&=\ii\hbar (\alpha+\beta)z_1\D z_2, & [z_3^*,\D
z_4^*]&=\ii\hbar(\alpha+\beta)z_1^* \D z_2^*,\\
[z_4,\D z_3]&=-\ii\hbar (\alpha+\beta)z_2\D z_1, & [z_4^*,\D
z_3^*]&=-\ii\hbar(\alpha+\beta)z_2^*\D z_1^*,\\
[z_3,\D z_3^*]&=\ii\hbar \alpha z_1\D z_1^*-\ii\hbar \beta z_2 \D
z_2^*, & [z_4,\D z_4^*]&=\ii\hbar \beta z_1\D z_1^*-\ii\hbar \alpha
z_2\D z_2^*,
\end{align*}
and the anti-commutation relations between one-forms
\begin{align*}
\{\D z_3,\D z_4\}&=\ii\hbar(\alpha+\beta) \D z_1\D z_2, & \{\D
z_3^*,\D
z_4^*\}&=\ii\hbar(\alpha+\beta) \D z_1^* \D z_2^*,\\
\{\D z_3,\D z_3^*\}&=\ii\hbar \alpha\D z_1\D z_1^*-\ii\hbar \beta\D
z_2 \D z_2^*, & \{\D z_4,\D z_4^*\}&=\ii\hbar \beta \D z_1\D
z_1^*-\ii\hbar \alpha\D z_2\D z_2^*,
\end{align*}
with all other relations undeformed.
\end{lem}

\proof One views the classical calculus $\Omega(\C^4)$ as a left
$H$-comodule algebra and accordingly computes the deformed product
using the twisting cocycle $F$. Since the exterior derivative $\D$
commutes with the $H$-coaction \eqref{tr-co1}, it is straightforward
to observe that the (anti-)commutation relations in the deformed
calculus $\Omega(\C^4_\hbar)$ are just the same as the algebra
relations in $\A[\C^4_\hbar]$ but with $\D$ inserted
appropriately.\endproof

\begin{lem}\label{le:stcalc}
The twisted differential calculus $\Omega(\RR^4_\hbar)$ is generated
by the degree zero elements $\zeta_1,\zeta_1^*,\zeta_2,\zeta_2^*$
and the degree one elements
$\D\zeta_1,\D\zeta_1^*,\D\zeta_2,\D\zeta_2^*$. The relations in the
calculus are not deformed.
\end{lem}

\proof Once again, the classical calculus $\Omega(\RR^4)$ is
deformed as a twisted left $H$-comodule algebra. Although the
products of functions and differential forms in the calculus are
indeed twisted, one finds that the extra terms which appear in the
twisted product all vanish in the expressions for the
(anti-)commutators ({\em cf.} \cite{bm:qtt} for full
details).\endproof

In particular, we see that the vector space $\Omega^2(\RR^4_\hbar)$
is the same as it is classically. Since the coaction of $H$ on
$\A[\RR^4]$ is by isometries, the Hodge $*$-operator
$*:\Omega^2(\RR^4)\to\Omega^2(\RR^4)$ commutes with the $H$-coaction
in the sense that
$$
\Delta_\pi(*\omega)=(\id\otimes *)\Delta_\pi(\omega),\qquad
\omega\in\Omega^2(\RR^4),
$$
so there is also a Hodge operator
$*_\hbar:\Omega^2(\RR^4_\hbar)\to\Omega^2(\RR^4_\hbar)$ defined by
the same formula as in the classical case. In particular, this means
that the decomposition of $\Omega^2(\RR^4_\hbar)$ into self-dual and
anti-self-dual two-forms,
$$
\Omega^2(\RR^4_\hbar)=\Omega^2_+(\RR^4_\hbar)\oplus\Omega^2_-(\RR^4_\hbar),
$$
is identical at the level of vector spaces to the corresponding
decomposition in the classical case.

The above lemmata are really just special cases of the cocycle
twisting procedure; recall that in fact our
`quantisation map' applies to every suitable $H$-covariant construction, in
particular to the coordinate algebra $\A[{\sf M}_k]$ of the space of
self-conjugate monads. We view $\A[{\sf M}_k]$ as a left
$H$-comodule algebra according to Lemma~\ref{le:mor} and write
$\A[{\sf M}_{k;\hbar}]$ for the corresponding cocycle-twisted
$H_F$-comodule algebra.

\begin{prop}\label{pr:moyal-mon}
The coordinate $*$-algebra $\A[{\sf M}_{k;\hbar}]$ is generated by
the matrix elements $M^j_{ab}$, $N^l_{dc}$ for $a,c=1,\ldots,k$ and
$b,d=1,\ldots,2k+2$, modulo the relations
\begin{align*}
[M^1_{ab},M^2_{rs}]&=\ii\hbar (\alpha-\beta)M^3_{ab}M^4_{rs}, &
[M^1_{ab},M^1_{rs}{}^*]&=\ii\hbar
\alpha M^3_{ab}M^3_{rs}{}^*+\ii\hbar \beta M^4_{ab}M^4_{rs}{}^*,\\
[M^1_{ab}{}^*,M^2_{rs}{}^*]&=\ii\hbar
(\alpha-\beta)M^3_{ab}{}^*M^4_{rs}{}^*, &
[M^2_{ab},M^2_{rs}{}^*]&=-\ii\hbar \beta
M^3_{ab}M^3_{rs}{}^*-\ii\hbar \alpha M^4_{ab}M^4_{rs}{}^*
\end{align*}
and the $*$-structure \eqref{monad star}. The generators $M^3$,
$M^4$, $M^3{}^*$, $M^4{}^*$ are central in the algebra.
\end{prop}

\proof From Lemma~\ref{le:mor} we read off the $H$-coaction on
generators $M^j$, $j=1,\ldots,4$, obtaining
\begin{align*}
M^1&\mapsto 1\otimes M^1-t_1^*\otimes M^3+t_2\otimes M^4, &
M^3&\mapsto 1\otimes M^3,\\
M^2&\mapsto 1\otimes M^2-t_2^*\otimes M^3-t_1\otimes M^4, &
M^4&\mapsto 1\otimes M^4,
\end{align*}
which we extend as a $*$-algebra map. The deformed relations follow
immediately from an application of the twisting formula
\eqref{com-prod}. The coaction of $H$ on $\A[{\sf M}_{k}]$ does not
depend on the matrix indices of the generators $M^j$, $N^l$,
$j,l=1,\ldots,4$, hence neither do the twisted commutation
relations. Similar computations yield the other commutation
relations as stated. In terms of the deformed product, the relations
\eqref{adhm eqs} are twisted into the relations
$$
\sum_r N^j_{dr} M^l_{rb}+ N^l_{dr} M^j_{rb}+\ii\hbar(\alpha+\beta)(\delta^{j1}\delta^{l2}-\delta^{j2}\delta^{l1})=0
$$
for each $b,d=1,\ldots, k$, where $\delta^{rs}$ is the Kronecker
delta symbol.\endproof

We think of $\A[{\sf M}_{k;\hbar}]$ as the coordinate algebra of a
noncommutative space ${\sf M}_{k;\hbar}$ of monads on $\C^4_\hbar$.
Although we do not have as many evaluation maps on $\A[{\sf
M}_{k;\hbar}]$ as we did in the classical case, we can nevertheless
work with the whole family ${\sf M}_{k;\hbar}$ at once.

\subsection{The construction of instantons on $\RR^4_\hbar$}
\label{se:moyaladhm} 

From the noncommutative space of monads
$\M_{k;\hbar}$ we may proceed as in \S\ref{se:adhmcon} to construct
families of instantons, now on the Moyal space $\RR^4_\hbar$.

Let $\widehat\RR^4$ denote the Pontryagin dual to the additive group
$\RR^4$ used in Eq.~\eqref{defhopf}. Given a pair of complex numbers
$\vec c:=(c_1,c_2)\in\C^2\simeq\widehat\RR^4$ we define unitary
elements $\vec u=(u_1,u_2)$ of the algebra $H_F$ by
\begin{equation}\label{units}
u_1=\textrm{exp}(\ii(c_1t_1+c_1^*t_1^*)),\quad
u_2=\textrm{exp}(\ii(c_2t_2+c_2^*t_2^*)).
\end{equation}
It is straightforward to check that $u_1$, $u_2$ are group-like
elements of (the smooth completion of) the Hopf algebra $H_F$, {\em
i.e.} they transform as $\Delta(u_j)=u_j\otimes u_j$ under the
coproduct $\Delta:H_F\to H_F\otimes H_F$.

\begin{lem}
There is a canonical left action of $H_F$ on the algebra
$\A[{\sfM}_{k;\hbar}]$ given by
\begin{align*}
u_1\tr M^1&=M^1-\hbar\alpha c_1 M^3, & u_1\tr M^2&=M^2+\hbar\alpha
c_1^*M^4,\\
u_1\tr M^1{}^*&=M^1{}^*-\hbar\alpha c_1^* M^3{}^*, & u_1\tr
M^2{}^*&=M^2{}^*+\hbar\alpha
c_1M^4{}^*,\\
u_2\tr M^1&=M^1+\hbar\beta c_2^*M^4, & u_2\tr M^2&=M^2+\hbar\beta
c_2
M^3,\\
u_2\tr M^1{}^*&=M^1{}^*+\hbar\beta c_2M^4{}^*, & u_2\tr
M^2{}^*&=M^2+\hbar\beta c_2^* M^3{}^*,
\end{align*}
with $u_j\tr M^l=M^l$ and $u_j\tr M^l{}^*=M^l{}^*$ for $l=3,4$.
\end{lem}

\proof Recall from the proof of Proposition~\ref{pr:moyal-mon} that
$\A[{\sfM}_{k;\hbar}]$ is a left $H_F$-comodule algebra; it is
therefore also a left $H_F$-module algebra according to the formula
\eqref{leftact}. Evaluating the $R$-matrix by expanding the
exponentials as power series, one finds that
\begin{align*}
\mathcal{R}(t_1,u_1)&=\mathcal{R}(t_1,\ii c_1^*t_1^*)=-\hbar\alpha
c_1^*, &
\mathcal{R}(t_1^*,u_1)&=\mathcal{R}(t_1^*,\ii c_1t_1)=\hbar\alpha c_1, \\
\mathcal{R}(t_2,u_2)&=\mathcal{R}(t_1,\ii c_2^*t_2^*)=-\hbar\beta
c_2^*, & \mathcal{R}(t_2^*,u_2)&=\mathcal{R}(t_2^*,\ii c_2
t_2)=\hbar\beta c_2,
\end{align*}
with all other combinations evaluating as zero. Using the fact that
the unitaries $u_j$ are group-like elements of the Hopf algebra
$H_F$, one finds the actions to be as stated.\endproof

In turn, there is an infinitesimal version of the $H_F$-action on
$\A[{\sfM}_{k;\hbar}]$, given by
\begin{align*}
t_1\tr M^1&=\ii\hbar\alpha M^3, & t_1^*\tr M^1{}^*&=-\ii\hbar\alpha
M^3{}^*, & t_1^*\tr M^2&=-\ii\hbar\alpha M^4, & t_1\tr
M^2{}^*&=\ii\hbar\alpha
M^4{}^*,\\
t_2^*\tr M^1&=-\ii\hbar\beta M^4, & t_2\tr M^1{}^*&=\ii\hbar\beta
M^4{}^*, & t_2\tr M^2&=-\ii\hbar\beta M^3, & t_2^*\tr
M^2{}^*&=\ii\hbar\beta M^3{}^*,
\end{align*}
with $t_j\tr M^l=0$ and $t_j\tr M^l{}^*=0$ for all other possible
combinations of generators. Either way, we obtain a group action
$$
\gamma:\widehat\RR^4\to\mathrm{Aut}\,\A[{\sfM}_{k;\hbar}]
$$
of the Pontryagin dual $\widehat\RR^4$ on the coordinate algebra
$\A[{\sfM}_{k;\hbar}]$ by $*$-automorphisms. We also
form the smash product algebra $\A[{\sf M}_{k;\hbar}]\lcross H_F$ associated to the above $H_F$-action,
whose multiplication is defined by the formula \eqref{cross}. With
the coproduct $\Delta(t_j)=1\otimes t_j+t_j\otimes 1$ on $H_F$, we
find in particular the formula
$$
(M^j_{ab}\otimes t_r)(M^l_{cd}\otimes t_s)=M^j_{ab}M^l_{cd}\otimes
t_rt_s+M^j_{ab}(t_r\tr M^l_{cd})\otimes t_s,
$$
with similar expressions for products involving the conjugate
generators $M^j{}^*$. The corresponding algebra relations between
such elements are given by
\begin{equation}\label{hbar cross}
[M^j_{ab}\otimes t_r,M^l_{cd}\otimes t_s]=[M^j_{ab},M^l_{cd}]\otimes
t_rt_s+M^j_{ab}(t_r\tr M^l_{cd})\otimes t_s-M^l_{cd}(t_s\tr
M^j_{ab})\otimes t_r
\end{equation}
for $j,l=1,\ldots,4$ and $r,s=1,2$, with similar formul{\ae}
occurring when the generators $M^j$ and $t_r$ are replaced by their
conjugates. Of course, these relations are just a small part of the
full algebra structure in the smash product $\A[{\sf
M}_{k;\hbar}]\lcross H_F$, but those in Eq.~\eqref{hbar cross} are
the ones that we will need later on in the paper.

\begin{rem}\textup{This situation is a special case of Example~\ref{ex:smash}. Recall
that, upon making suitable completions of our algebras, we can think
of the smash product algebra 
$$
\A[{\sf M}_{k;\hbar}]\lcross H_F=\A[{\sf M}_{k;\hbar}]\lcross \A[\RR^4]
$$ 
as being equivalent to
the crossed product algebra $\A[{\sf M}_{k;\hbar}]\lcross_\gamma
\,\widehat\RR^4$.}\end{rem}

Thanks to the functorial nature of the cocycle twisting, {\em
mutatis mutandis} the ADHM construction goes through as described in
\S\ref{se:adhmcon}. In the following, we highlight the main
differences which arise as a consequence of the quantisation
procedure. The next lemma takes care of an important technical
point: as well as twisting the relations in the algebras
$\A[{\sfM}_{k}]$ and $\A[\C^4]$, we also have to deform the
cross-relations in the tensor product algebra
$\A[{\sfM}_{k}]\otimes\A[\C^4]$.

\begin{lem}
The algebra structure of the twisted tensor product algebra
$\A[{\sfM}_{k;\hbar}]\utimes\A[\C^4_\hbar]$ is given by the
relations in the respective subalgebras $\A[{\sfM}_{k;\hbar}]$ and
$\A[\C^4_\hbar]$ determined above, together with the cross-relations
\begin{align*}
z_3M^1&=M^1z_3-\ii\hbar\beta M^4z_2, & z_3M^2&=M^2
z_3-\ii\hbar\alpha M^4z_1,\\
z_4M^1&=M^1z_4+\ii\hbar\alpha M^3z_2, & z_4M^2&=M^2z_4+\ii\hbar\beta
M^3z_1,\\
z_3^*M^1&=M^1z_3^*+\ii\hbar\alpha M^3z_1^*, &
z_3^*M^2&=M^2z_3^*-\ii\hbar\beta M^3z_2^*,\\
z_4^*M^1&=M^1z_4^*+\ii\hbar\beta M^4z_1^*, &
z_4^*M^2&=M^2z_4^*-\ii\hbar\alpha M^4z_2^*
\end{align*}
and their conjugates. The generators $z_1, z_2, M^3, M^4$ are
central.
\end{lem}

\proof The classical algebra $\A[{\sf M}_k]\otimes\A[\C^4]$ is a
left comodule $*$-algebra under the tensor product $H_F$-coaction
defined by Eq.~\eqref{mon}. The twisted product is determined by the
formula \eqref{com-prod}, with the non-trivial cross-terms in the
deformed algebra being the ones stated in the lemma. We denote the
deformed algebra by $\A[{\sfM}_{k;\hbar}]\utimes\A[\C^4_\hbar]$,
with the symbol $\utimes$ to remind us that the tensor product
algebra structure is not the usual one, but has been twisted as
well.\endproof

Just as in the classical situation, we have a pair of matrices
$\sigma_z$ and $\tau_z$,
\begin{align*}
\sigma_z&=\sum_j M^j\otimes z_j, & \tau_z&=\sum_j N^j\otimes z_j,
\end{align*}
whose entries this time live in the twisted algebra
$\A[{\sfM}_{k;\hbar}]\utimes\A[\C^4_\hbar]$. The resulting matrix
$\sfV:=\begin{pmatrix}\sigma_z&\sigma_{J(z)}\end{pmatrix}$ is a
$2k\times(2k+2)$ matrix with entries in
$\A[{\sfM}_{k;\hbar}]\utimes\A[\C^4_\hbar]$. We set
$\rho^2:=\sfV^*\sfV$. From the projection $\Qp:=\sfV\rho^{-2}\sfV^*$
we construct the complementary matrix $\Pp:=\mathbbm{1}_{2k+2}-\Qp$,
which has entries in the algebra
$\A[{\sfM}_{k;\hbar}]\utimes\A[\RR^4_\hbar]$.

It is clear that this matrix $\Pp$ is a self-adjoint idempotent,
$\Pp^2=\Pp=\Pp^*$, but it does not define an honest family of
projections in the sense of Definition~\ref{de:fam-bun}. Recall
that, to define such a family, we need a matrix with entries in an
algebra of the form $A\otimes \A[\RR^4_\hbar]$ for some `parameter algebra' $A$, whereas the quantisation procedure has produced a
projection $\Qp$ with entries in a {\em twisted} tensor product
$\A[{\sfM}_{k;\hbar}]\utimes\A[\RR^4_\hbar]$. We may nevertheless
recover a genuine family of projections using the following lemma,
in which we use the Sweedler notation $Z\mapsto Z\mo\otimes Z\bz$
for the left coaction $\A[\C^4_\hbar]\to H_F\otimes \A[\C^4_\hbar]$
defined in Eqs.~\eqref{tr-co1}--\eqref{tr-co2}.

\begin{lem}\label{le:beta}
There is a canonical $*$-algebra map
$$
\mu:\A[{\sfM}_{k;\hbar}]\utimes\A[\C^4_\hbar]\to\left(\A[{\sfM}_{k;\hbar}]\lcross
H_F\right)\otimes\A[\C^4_\hbar]
$$
defined by $\mu(M\otimes Z)=M\otimes Z\mo\otimes Z\bz$ for each
$M\in \A[{\sfM}_{k;\hbar}]$ and $Z\in \A[\C^4_\hbar]$.
\end{lem}

\proof This follows from a straightforward verification. One checks
that
\begin{align*}
\mu(M\otimes Z)\mu(M'\otimes Z')&=(M\otimes Z\mo\otimes
Z\bz)(M'\otimes Z'{}\mo\otimes Z'{}\bz) \\
&=M(Z\mo{}\o\tr M')\otimes Z\mo{}\t Z'\mo\otimes Z\bz Z'\bz\\
&=\mathcal{R}(M'{}\mo,Z\mo{}\o)MM'{}\bz\otimes Z\mo{}\t Z'\mo\otimes
Z\bz Z'\bz\\
&=\mathcal{R}(M'{}\mo,Z\mo)MM'{}\bz\otimes Z(Z\bz{}\mo)
Z'{}\mo\otimes
(Z\bz{}\bz) Z'{}\bz\\
&=\mu\left( \mathcal{R}(M'{}\mo,Z\mo) MM'{}\bz\otimes Z\bz
Z'\right)\\&=\mu\left( (M\otimes Z)(M'\otimes Z')\right)
\end{align*}
so that $\mu$ is an algebra map, as well as
\begin{align*}
\left(\mu(M\otimes Z)\right)^*&=(M\otimes Z\mo\otimes Z\bz)^*
=(M\otimes Z\mo)^*\otimes Z\bz{}^*\\
&=\mathcal{R}(M\mo{}^*,(Z\mo{}\o)^*)(M\bz{}^*\otimes
(Z\mo{}\t)^*)\otimes Z\bz{}^*\\
&=\mu\left( \mathcal{R}(M\mo{}^*,Z\mo{}^*)(M\bz{}^*\otimes
Z\bz{}^*)\right)\\&=\mu\left( (M\otimes Z)^*\right)
\end{align*}
so that $\mu$ respects the $*$-structures as well.\endproof

\begin{rem}\textup{Lemma~\ref{le:beta} is an example of Majid's `bosonisation' construction \cite{ma:book}, which converts noncommutative `braid statistics' (in our case described by the twisted tensor product $\utimes$) into commutative `ordinary statistics' (described by the usual tensor product $\otimes$).}
\end{rem}

As a consequence of Lemma~\ref{le:beta}, we find that there are maps
\begin{align}\label{tildesigma}
\tilde\sigma_z:\mH\otimes\A[\C^4_\hbar]&\to
\left(\A[{\sfM}_{k;\hbar}]\lcross H_F\right)\otimes
\mK\otimes\A[\C^4_\hbar],\\
\label{tildetau}\tilde\tau_z:\mK\otimes\A[\C^4_\hbar]&\to
\left(\A[{\sfM}_{k;\hbar}]\lcross H_F\right)\otimes
\mL\otimes\A[\C^4_\hbar],
\end{align}
defined by composing $\sigma_z$ and $\tau_z$ with the map $\mu$. Explicitly, these maps are given by
\begin{align*}
\tilde\sigma_z&:=\sum_r M^r\otimes z_r{}\mo\otimes z_r{}\bz, &
\tilde\tau_z&:=\sum_r N^r\otimes z_r{}\mo\otimes z_r{}\bz,
\end{align*}
which are respectively $k\times (2k+2)$ and $(2k+2)\times k$
matrices with entries in the noncommutative algebra
$(\A[{\sfM}_{k;\hbar}]\lcross H_F)\otimes\A[\C^4_\hbar]$. With this
in mind, we form the $(2k+2)\times 2k$ matrix
$\tsfV:=\begin{pmatrix}\tilde\sigma_z&\tilde\sigma_{J(z)}\end{pmatrix}$.

\begin{prop}
The $(2k+2)\times(2k+2)$ matrix $\tQp=\tsfV\trho^{-2}\tsfV^*$ is a
projection, $\tQp^2=\tQp=\tQp^*$, with entries in the algebra
$(\A[{\sfM}_{k;\hbar}]\lcross H_F)\otimes\A[\C^4_\hbar]$ and trace
equal to $2k$.
\end{prop}

\proof The fact that $\tQp$ is a projection follows from the fact
that $\Qp$ is a projection and that
$\mu:\A[{\sfM}_{k;\hbar}]\utimes\A[\C^4_\hbar]\to(\A[{\sfM}_{k;\hbar}]\lcross
H_F)\otimes\A[\C^4_\hbar]$ is a $*$-algebra map. By construction,
the entries of the matrix $\rho^2$ are central in the algebra
$(\A[{\sfM}_{k;\hbar}]\lcross H_F)\otimes\A[\C^4_\hbar]$ (this
follows from the fact that its matrix entries are coinvariant under
the left $H_F$-coaction), from which it follows that the trace
computation in Proposition~\ref{pr:adhm-pr} is valid in the
noncommutative case as well \cite{bl:mod}.\endproof

From the projection $\tQp$ we construct as before the complementary
projection $\tPp:=\mathbbm{1}_{2k+2}-\tQp$; it has entries in the
algebra $(\A[{\sfM}_{k;\hbar}]\lcross H_F)\otimes\A[\RR^4_\hbar]$
and has trace equal to two. In analogy with
Definition~\ref{de:fam-bun}, the finitely-generated projective
module
$$
\E:=\Pp\left((\A[{\sfM}_{k;\hbar}]\lcross
H_F)\otimes\A[\RR^4_\hbar]\right)^{2k+2}
$$
defines a family of rank two vector bundles over $\RR^4_\hbar$
parameterised by the noncommutative algebra
$\A[{\sfM}_{k;\hbar}]\lcross H_F$. We equip this family of vector
bundles with the family of Grassmann connections associated to the
projection $\Pp$.

\begin{prop}
The curvature $F=\Pp((\id\otimes \D)\Pp)^2$ of the Grassmann family
of connections $\n:=(\id\otimes \D)\circ\Pp$ is anti-self-dual.
\end{prop}

\proof From Lemma~\ref{le:stcalc} we know that the space of
two-forms $\Omega^2(\RR^4_\hbar)$ and the Hodge $*$-operator
$*_\hbar:\Omega^2(\RR^4_\hbar)\to\Omega^2(\RR^4_\hbar)$ are
undeformed and equal to their classical counterparts; similarly for
the decomposition
$\Omega^2(\RR^4_\hbar)=\Omega^2_+(\RR^4_\hbar)\oplus\Omega^2_-(\RR^4_\hbar)$
into self-dual and anti-self-dual two-forms. This identification of
the `quantum' with the `classical' spaces of two-forms survives the
tensoring with the parameter space $\A[{\sfM}_{k;\hbar}]\lcross
H_F$, which yields that $(\A[{\sfM}_{k;\hbar}]\lcross
H_F)\otimes\Omega^2_\pm(\RR^4_\hbar)$ and $(\A[{\sfM}_{k}]\otimes
H)\otimes\Omega^2_\pm(\RR^4)$ are isomorphic as vector spaces.
Computing the curvature $F$ in exactly the same way as in
Proposition~\ref{pr:asd}, we see that it must be anti-self-dual,
since the same is true in the classical case.\endproof

\subsection{The Moyal-deformed ADHM equations}\label{se:ns-adhm}
The noncommutative ADHM construction of the previous section
produced families of instantons on $\RR^4$ parameterised by the
noncommutative algebra $\A[{\sfM}_{k;\hbar}]\lcross H_F$. We
interpret the latter as an algebra of coordinate functions
on some underlying `quantum' parameter space, within which we shall seek a subspace of classical parameters. To this end,
we introduce elements of $\A[{\sfM}_{k;\hbar}]\lcross
H_F$ defined by
\begin{align*}
\widetilde M^1_{ab}:&=M^1_{ab}\otimes 1+M^3_{ab}\otimes \half
t_1^*-M^4_{ab}\otimes \half t_2, & \widetilde
M^3_{ab}:&=M^3_{ab}\otimes 1,\\ \widetilde
M^2_{ab}:&=M^2_{ab}\otimes 1+M^3_{ab}\otimes \half
t_2^*+M^4_{ab}\otimes \half t_1, & \widetilde
M^4_{ab}:&=M^4_{ab}\otimes 1
\end{align*}
for each $a=1,\ldots, k$ and $b=1,\ldots, 2k+2$, together with their
conjugates $\widetilde M^j_{ab}{}^*$, $j=1,\ldots,4$.

\begin{defn}We write $\A[\mathfrak{M}(k;\hbar)]$ for the subalgebra of
$\A[{\sfM}_{k;\hbar}]\lcross H_F$ generated by the elements
$\widetilde M^j_{ab}$, $\widetilde M^l_{dc}{}^*$, $j,l=1,\ldots,4$.
\end{defn}

\begin{prop}
The algebra $\A[\mathfrak{M}(k;\hbar)]$ is a commutative
$*$-subalgebra of the smash product $\A[{\sfM}_{k;\hbar}]\lcross
H_F$.
\end{prop}

\proof This follows from direct computation. The generators
$\widetilde M^3$ and $\widetilde M^4$ are clearly central. On the
other hand, we also have
\begin{align*}
\widetilde M^1\widetilde M^2&=M^1M^2\otimes 1+M^1M^3\otimes \half
t_2^*+M^1M^4\otimes \half t_1+M^3M^2\otimes \half t_1^*\\&
\qquad\qquad\qquad\qquad -M^4M^2\otimes \half
t_2-\half\ii\hbar\alpha M^3M^4\otimes
1+\half\ii\hbar\beta M^4M^3\otimes 1,\\
\widetilde M^2\widetilde M^1&=M^2M^1\otimes 1+M^3M^1\otimes \half
t_2^*+M^4M^1\otimes \half t_1+M^2M^3\otimes \half t_1^*\\&
\qquad\qquad\qquad\qquad -M^2M^4\otimes \half
t_2+\half\ii\hbar\alpha M^4M^3\otimes 1-\half\ii\hbar\beta
M^3M^4\otimes 1,
\end{align*}
from which it follows that the commutator is given by
$$
[\widetilde M^1,\widetilde M^2]=[M^1,M^2]\otimes
1-\ii\hbar(\alpha-\beta)M^3M^4\otimes 1=0.
$$
All other commutators are shown to vanish in the same way.\endproof

Although we have made a change of generators, this does not affect
the family of instantons constructed in the previous section. In
order to show this, let
$$
\tr':H_F\otimes \A[\mathfrak{M}(k;\hbar)]\to
\A[\mathfrak{M}(k;\hbar)]
$$
be the left action of $H_F$ on $\A[\mathfrak{M}(k;\hbar)]$ defined
on generators by
\begin{align*}
t_1\tr' \widetilde M^1&=\ii\hbar\alpha \widetilde M^3, &
t_1^*\tr' \widetilde M^1{}^*&=-\ii\hbar\alpha \widetilde M^3{}^*, \\
t_1^*\tr' \widetilde M^2&=-\ii\hbar\alpha \widetilde M^4, & t_1\tr'
\widetilde M^2{}^*&=\ii\hbar\alpha
\widetilde M^4{}^*,\\
t_2^*\tr' \widetilde M^1&=-\ii\hbar\beta \widetilde M^4, & t_2\tr'
\widetilde M^1{}^*&=\ii\beta \widetilde M^4{}^*,
\\ t_2\tr' \widetilde M^2&=-\ii\hbar\beta \widetilde M^3, & t_2^*\tr'
\widetilde M^2{}^*&=\ii\hbar\beta \widetilde M^3{}^*,
\end{align*}
together with $t_j\tr' \widetilde M^l=0$ and $t_j\tr' \widetilde
M^l{}^*=0$ for $l=3,4$. Let us write
$\A[\mathfrak{M}(k;\hbar)]\lcross H_F$ for the smash product algebra
associated to the action $\tr'$.

\begin{thm}
There is a $*$-algebra isomorphism
$\phi:\A[\mathfrak{M}(k;\hbar)]\lcross H_F\to
\A[{\sfM}_{k;\hbar}]\lcross H_F$ defined for each $h\in H_F$ by
\begin{align*}
\widetilde M^1_{ab}\otimes h&\mapsto M^1_{ab}\otimes
h+M^3_{ab}\otimes \half t_1^*h-M^4_{ab}\otimes \half t_2h, &
\widetilde
M^3_{ab}\otimes h&\mapsto M^3_{ab}\otimes h, \\
\widetilde M^2_{ab}\otimes h&\mapsto M^2_{ab}\otimes
h+M^3_{ab}\otimes \half t_2^*h-M^4_{ab}\otimes \half t_1h, &
\widetilde M^4_{ab}\otimes h&\mapsto M^4_{ab}\otimes h
\end{align*}
and extended as a $*$-algebra map.
\end{thm}

\proof It is clear that this map is an isomorphism of vector spaces
with inverse
\begin{align*}
M^1_{ab}\otimes h&\mapsto \widetilde M^1_{ab}\otimes h-\widetilde
M^3_{ab}\otimes \half t_1^*h+\widetilde M^4_{ab}\otimes \half t_2h,
& M^3_{ab}\otimes h&\mapsto \widetilde M^3_{ab}\otimes h, \\
M^2_{ab}\otimes h&\mapsto \widetilde M^2_{ab}\otimes h-\widetilde
M^3_{ab}\otimes \half t_2^*h-\widetilde M^4_{ab}\otimes \half t_1h,
& M^4_{ab}\otimes h&\mapsto \widetilde M^4_{ab}\otimes h.
\end{align*}
By definition, the map $\phi$ is a $*$-algebra homomorphism on the
subalgebra $\A[\mathfrak{M}(k;\hbar)]$, so we just have to check
that it preserves the cross-relations between
$\A[\mathfrak{M}(k;\hbar)]$ and the subalgebra $H_F$. This is
straightforward to verify: one has for example that
\begin{align*}
\phi(1\otimes t_1)\phi(\widetilde M^1\otimes 1)&=(1\otimes
t_1)(M^1_{ab}\otimes 1+M^3_{ab}\otimes \half t_1^*-M^4_{ab}\otimes
\half t_2)\\&=(M^1_{ab}\otimes t_1+M^3_{ab}\otimes \half
t_1^*t_1-M^4_{ab}\otimes \half t_2t_1)+(t_1\tr M^1)\otimes
1\\&=\phi(\widetilde M^1\otimes t_1+(t_1\tr'\widetilde M^1)\otimes
1)\\&=\phi((1\otimes t_1)(\widetilde M^1\otimes 1)).
\end{align*}
The remaining relations are verified in the same way.\endproof

Our goal is now to see that the parameters corresponding to the
subalgebra $H_F$ can be removed and that there is a family of
instantons parameterised by the commutative algebra
$\A[\mathfrak{M}(k;\hbar)]$. This follows from the fact that there
is a right coaction
\begin{equation}\label{gauge-co}
\delta_R:\A[\mathfrak{M}(k;\hbar)]\lcross H_F\to
\left(A[\mathfrak{M}(k;\hbar)]\lcross H_F\right)\otimes H_F,\qquad
\delta_R:=\id\otimes\Delta,
\end{equation}
where $\Delta:H_F\to H_F\otimes H_F$ is the coproduct on the Hopf
algebra $H_F$. This coaction is by `gauge transformations', in the
sense that the projections $\widetilde\Pp\otimes 1$ and
$\delta_R(\widetilde\Pp)$ are unitarily equivalent in the matrix
algebra $\M_{2k+2}\left((\A[\mathfrak{M}(k;\hbar)]\lcross
H_F)\otimes H_F\right)$ and so they define gauge equivalent families
of instantons \cite{bl:mod}. This means that the parameters
determined by the subalgebra $H_F$ in
$\A[\mathfrak{M}(k;\hbar)]\lcross H_F$ are just gauge parameters and
so they may be removed. Indeed, by passing to the subalgebra of
$\A[\mathfrak{M}(k;\hbar)]\lcross H_F$ consisting of coinvariant
elements under the coaction \eqref{gauge-co}, {\em viz}.
$$
\A[\mathfrak{M}(k;\hbar)]\cong\{a\in\A[\mathfrak{M}(k;\hbar)]\lcross
H_F~|~\delta_R(a)=a\otimes 1\},
$$
we obtain a projection $\Pp_{k;\hbar}$ with entries in
$\A[\mathfrak{M}(k;\hbar)]\otimes\A[\RR^4_\hbar]$. The precise
construction of the projection $\Pp_{k;\hbar}$ goes exactly as in
\cite{bl:mod}, as does the proof of the fact that the Grassmann
family of connections $\n=\Pp_{k;\hbar}\circ(\id\otimes \D)$ has
anti-self-dual curvature and hence defines a family of instantons on
$\RR^4_\hbar$.

For each point $x\in\mathfrak{M}(k;\hbar)$ there is an evaluation
map
$$
\textrm{ev}_x\otimes \id:\A[\mathfrak{M}(k;\hbar)]\otimes
\A[\C^4_\hbar]\to \A[\C^4_\hbar],
$$
which in turn defines a self-conjugate monad over the noncommutative
space $\C^4_\hbar$. In analogy with Remark~\ref{re:framing}, the
matrices $(\textrm{ev}_x\otimes\id)\widetilde \sigma_z$ and
$(\textrm{ev}_x\otimes\id)\widetilde \tau_z$ determine a complex of
free right $\A[\C^4_\hbar]$-modules
\begin{equation}\label{monad-hbar}
0\to \mH\otimes \A[\C^4_\hbar]
\xrightarrow{(\textrm{ev}_x\otimes\id)\tilde\sigma_z} \mK\otimes
\A[\C^4_\hbar] \xrightarrow{(\textrm{ev}_x\otimes\id)\tilde\tau_z}
\mL\otimes \A[\C^4_\hbar]\to 0.
\end{equation}
The same evaluation map determines a projection
$(\textrm{ev}_x\otimes\id)\Pp_{k;\hbar}$ and hence an instanton
connection on $\RR^4_\hbar$.

As described in Proposition~\ref{pr:gauge}, the gauge freedom in the
classical ADHM construction is precisely the freedom determined by
the choice of bases of the vector spaces $\mH,\mK,\mL$. Clearly we
also have this freedom in the noncommutative construction as well:
 we write $\sim$ for the equivalence relation induced on the space $\mathfrak{M}(k;\hbar)$ by such changes of basis ({\em cf}. Definition~\ref{de:mon-eq}). This leads to the following explicit description of the parameter space
$\mathfrak{M}(k;\hbar)$ ({\em cf}. \cite{kko}).

\begin{thm}\label{th:moy-adhm}
For each positive integer $k\in \ZZ$, the space
$\mathfrak{M}(k;\hbar)/\sim$ of equivalence classes of self-conjugate monads over $\C^4_\hbar$
is the quotient of the set of complex matrices $B_1,B_2\in\M_k(\C)$,
$I\in \M_{2\times k}(\C)$, $J\in\M_{k\times 2}(\C)$ satisfying the
equations
\begin{enumerate}[\hspace{0.5cm}(i)]
\item $[B_1,B_2]+IJ=0$,
\item
$[B_1,B_1^*]+[B_2,B_2^*]+II^*-J^*J=-\ii\hbar(\alpha+\beta)\mathbbm{1}_{k}$
\end{enumerate}
by the action of $\U(k)$ given by
\begin{equation*}
B_1\mapsto gB_1g^{-1},\qquad B_2\mapsto gB_2g^{-1}, \qquad I\mapsto
gI,\qquad J\mapsto Jg^{-1}
\end{equation*}
for each $g\in\U(k)$.
\end{thm}

\proof Recall that we write the monad maps $\tilde\sigma_z$,
$\tilde\tau_z$ as
$$
\tilde\sigma_z=\widetilde M^1z_1+\widetilde M^2z_2+\widetilde
M^3z_3+\widetilde M^4z_4,\qquad \tilde\tau_z=\widetilde
N^1z_1+\widetilde N^2z_2+\widetilde N^3z_3+\widetilde N^4z_4
$$
for constant matrices $\widetilde M^j$, $\widetilde N^l$, where
$j,l=1,\ldots,4$. Upon expanding out the condition
$\tilde\tau_z\circ\tilde\sigma_z=0$ and using the commutation
relations in Lemma~\ref{le:stdef}, we find the conditions
\begin{equation}\label{exp-mon1}
\widetilde N^j\widetilde M^l+\widetilde N^l\widetilde
M^j+\ii\hbar(\alpha+\beta)(\delta^{j1}\delta^{l2}-\delta^{j2}\delta^{l1})=0
\end{equation}
for $j,l=1,\ldots,4$. Recall from Lemma~\ref{le:triv} that the
typical fibre $\CP^1$ of the twistor fibration $\RR^4\times\CP^1$
has homogeneous coordinates $z_1$, $z_1^*$, $z_2$, $z_2^*$; it
follows that the `line at infinity' $\ell_\infty$ is recovered by
setting $z_1=z_2=0$. On this line, the monad condition
$\tilde\tau_z\circ\tilde\sigma_z=0$ becomes
\begin{equation}\label{rest}
\widetilde N^3\widetilde M^4+\widetilde N^4\widetilde M^3=0, \qquad
\widetilde N^3\widetilde M^3=0,\qquad \widetilde N^4\widetilde
M^4=0.
\end{equation}
Moreover, when $z_1=z_2=0$ we see from the relations
\eqref{def-h1}--\eqref{def-h2} that the coordinates $z_3$, $z_4$ and
their conjugates are mutually commuting, so that the line
$\ell_\infty$ is classical. The self-conjugacy of the monad implies
that the restricted bundle over $\ell_\infty$ is trivial; therefore
we can argue as in \cite{oss:vb} to show that the map $\widetilde
N^3\widetilde M^4=-\widetilde N^4\widetilde M^3$ is an isomorphism.
Using these conditions we choose bases for $\mH,\mK,\mL$ such that
$\widetilde N^3\widetilde M^4=\mathbbm{1}_k$ and
\begin{align*}
\widetilde M^3&=\begin{pmatrix}\mathbbm{1}_{k\times k} \\ 0_{k\times k} \\
0_{2\times k}\end{pmatrix}, & \widetilde M^4&=\begin{pmatrix}0_{k\times k} \\\mathbbm{1}_{k\times k}\\
0_{2\times k}\end{pmatrix}, & \widetilde
N^3&=\begin{pmatrix}0_{k\times k} \\ \mathbbm{1}_{k\times k} \\
0_{k\times 2}\end{pmatrix}^{\textrm{tr}}, & \widetilde
N^4&=\begin{pmatrix}-\mathbbm{1}_{k\times k} \\ 0_{k\times k} \\
0_{k\times 2}\end{pmatrix}^{\textrm{tr}}.
\end{align*}
Now invoking conditions \eqref{exp-mon1} for $j=3,4$ and $l=1,2$,
the remaining matrices are necessarily of the form
\begin{align*}
\widetilde M^1&=\begin{pmatrix}B_1\\B_2\\J\end{pmatrix}, &
\widetilde M^2&=\begin{pmatrix}B_1'\\B_2'\\J'\end{pmatrix}, &
\widetilde N^1&=\begin{pmatrix}-B_2
\\ B_1 \\ I\end{pmatrix}^{\textrm{tr}}, & \widetilde
N^2&=\begin{pmatrix}-B_2' \\ B_1' \\ I'\end{pmatrix}^{\textrm{tr}}.
\end{align*}
Using the conditions $\tilde\tau_{J(z)}^*=-\tilde\sigma_z$ and
$\tilde\sigma_{J(z)}^*=\tilde\tau_z$, which correspond to the
requirement that the monad be self-conjugate, we find that
$$
B_1'=-B_2^*,\qquad B_2'=B_1^*,\qquad J'=I^*,\qquad I'=-J^*.
$$
Thus in order to satisfy the condition
$\tilde\tau_z\circ\tilde\sigma_z=0$ it remains only to impose the
conditions \eqref{exp-mon1} in the cases $j=l=1$ and $j=1$, $l=2$.
The first of these is condition (i) in the theorem; the second case
is equivalent to requiring
$$
[B_1,B_1^*]+[B_2,B_2^*]+II^*-J^*J+\ii\hbar(\alpha+\beta)=0,
$$
giving condition (ii) in the theorem. Just as in the classical case
\cite{don}, it is evident that the remaining gauge freedom in this
calculation is given by the stated action of $\U(k)$, whence the
result.
\endproof

Finally, we comment on a significant difference between the
parameter space ${\sfM}_k$ of instantons on the classical space
$\RR^4$ and the parameter space $\mathfrak{M}(k;\hbar)$ of
instantons on the Moyal-deformed version $\RR^4_\hbar$. Recall that,
in the classical ADHM construction of \S\ref{se:adhmcon}, we needed
to assume that the algebra-valued matrix $\rho^2$ of Eq.~\eqref{rho}
is invertible. Formally adjoining an inverse $\rho^{-2}$ to the
algebra $\M_k(\C)\otimes \A[{\sf M}_{k}]\otimes \A[\RR^4]$ resulted
in the deletion of a collection of points from the parameter space
${\sf M}_{k}$. In contrast, this noncommutative ADHM construction
does not require this. Using the Moyal ADHM equations themselves one shows that the  matrix $\rho^2$, which now having passed to the commutative parameter space has entries in the algebra $\A[\mathfrak{M}(k;\hbar)]\otimes\A[\RR^4_\hbar]$, is automatically invertible (we refer to \cite{ns:ins,fur} for a proof).

\section{The Connes-Landi Noncommutative Plane
$\RR^4_\theta$}\label{se:toric} 

Next we turn to the construction of instantons on the noncommutative plane $\RR^4_\theta$, which is an example of a toric noncommutative manifold (or isospectral deformation) in the sense of \cite{cl:id}. In particular, $\RR^4_\theta$ is obtained as a localisation of the Connes-Landi quantum four-sphere $S^4_\theta$, just as in Lemma~\ref{st-proj} ({\em cf}. \cite{lvs:teh}), although here we shall obtain it directly from classical $\RR^4$ by cocycle twisting.

\subsection{Toric deformation of the space of monads}
Whereas the Moyal space-time $\RR^4_\hbar$ was obtained by
cocycle twisting along an action of the group of translation
sysmmetries of space-time, the noncommutative
space-time $\RR^4_\theta$ is constructed by deforming the classical coordinate
algebra $\A[\RR^4]$ along an action of a group of rotational
symmetries.

Indeed, for the twisting Hopf algebra we take $H=\A[\TT^2]$, the
algebra of coordinate functions on the two-torus $\TT^2$. It is the
commutative unital algebra
$$
\A[\TT^2]:=\A[s_j,s_j^{-1}~|~j=1,2]
$$
equipped with the Hopf $*$-algebra structure
\begin{equation}\label{hopf-tor}
s_j^*=s_j^{-1},\qquad \Delta(s_j)=s_j\otimes s_j,\qquad
\ep(s_j)=1,\qquad S(s_j)=s_j^{-1}
\end{equation}
for $j=1,2$, with $\Delta$, $\ep$ extended as $*$-algebra maps and
$S$ extended as a $*$-anti-algebra map.

In order to deform the twistor fibration, we need to equip the
various coordinate algebras with left $H$-comodule structures. There
is a Hopf algebra projection from $\A[\GL^+(2,\HH)]$ onto $H$,
defined on generators by
\begin{equation}\label{cl-proj}
\pi:\A[\GL^+(2,\HH)]\to H,\qquad  \begin{pmatrix}\alpha_1&-\alpha_2^*&\beta_1&-\beta_2^*\\\alpha_2&\alpha_1^*&\beta_2&\beta_1^*\\
\gamma_1&-\gamma_2^*&\delta_1&-\delta_2^*\\
\gamma_2&\gamma_1^*&\delta_2&\delta_1^*\end{pmatrix}\mapsto \begin{pmatrix}s_1&0&0&0\\0&s_1^*&0&0\\
0&0&s_2&0\\0&0&0&s_2^*\end{pmatrix}
\end{equation}
and extended as a $*$-algebra map. Using Eq.~\eqref{proj-co}, this
projection determines a left $H$-coaction $\Delta_\pi:\A[\C^4]\to
H\otimes \A[\C^4]$ by
\begin{equation}\label{H-coact2}
\A[\C^4]\to H\otimes \A[\C^4],\qquad z_j\mapsto \varsigma_j\otimes
z_j,
\end{equation}
extended as a $*$-algebra map, where we use the shorthand notation
$(\varsigma_j)=(s_1,s_1^*,s_2,s_2^*)$ for the generators of $H$.
Using the identification of generators in Eq.~\eqref{ch-inv}, this
induces a coaction on the space-time algebra,
\begin{equation}\label{sph-co2}
\A[\RR^4]\to H\otimes\A[\RR^4],\qquad \zeta_1\mapsto
\varsigma_1\varsigma_4\otimes \zeta_1,\quad \zeta_2\mapsto
\varsigma_2\varsigma_4\otimes \zeta_2,
\end{equation}
and extended as a $*$-algebra map, making $\A[\RR^4]$ into a left
$H$-comodule $*$-algebra.

As a twisting cocycle on $H$, we take the linear map defined on
generators by
\begin{equation}\label{tor-cocy}
F:H\otimes H\to \C,\qquad F(s_j,s_l)=\textup{exp}(\ii\pi\Theta_{jl})
\end{equation}
and extended as a Hopf bicharacter in the sense of Eq.\eqref{hopf-bic}. Here the deformation matrix
$\Theta$ is the $2\times 2$ real anti-symmetric matrix
$$
\Theta=(\Theta_{jl})=\tfrac{1}{2}\begin{pmatrix}0&\theta\\
-\theta&0\end{pmatrix}
$$
for $0<\theta<1$ a real parameter. It is straightforward to check
using the formul{\ae} \eqref{twprod}--\eqref{twanti} and
\eqref{tw-star} that the product, antipode and $*$-structure on $H$
are in fact undeformed by $F$, so that $H=H_F$ as a Hopf
$*$-algebra. However, the effect of the twisting on the $H$-comodule
algebras $\A[\C^4]$ and $\A[\RR^4]$ is non-trivial. In what follows
we write $\eta_{jl}:=F^{-2}(\varsigma_j,\varsigma_l)$, namely
\begin{equation}\label{eqn eta matrix}
(\eta_{jl})=\begin{pmatrix}1&1&\mu&\bar\mu\\1&1&\bar\mu&\mu\\\bar\mu&\mu&1&1\\\mu&\bar\mu&1&1\end{pmatrix},
\qquad \mu=e^{\ii\pi\theta}.
\end{equation}

\begin{lem}\label{le:sttheta}
The relations in the $H$-comodule algebra $\A[\C^4]$ are twisted
into
\begin{equation}\label{thetarels}
z_jz_l=\eta_{lj}z_lz_j,\qquad z_jz_l^*=\eta_{jl}z_l^*z_j,\qquad
z_j^*z_l=\eta_{jl}z_lz_j^*,\qquad z_j^*z_l^*=\eta_{lj}z_l^*z_j^*
\end{equation}
for each $j,l=1,\ldots,4$.
\end{lem}

\proof The cocycle-twisted product on the $H$-comodule algebra
$\A[\C^4]$ is defined by the formula \eqref{com-prod}. Just as in
Lemma~\ref{le:stdef}, the corresponding algebra relations can be
expressed using the $R$-matrix \eqref{R-braid}: in this case one
finds that the $R$-matrix takes the values
\begin{equation}\label{eval2}
\mathcal{R}(\varsigma_j,\varsigma_l)=F^{-2}(\varsigma_j,\varsigma_l)=\eta_{jl},\qquad
\mathcal{R}(\varsigma_j,\varsigma_l^*)=F^{-2}(\varsigma_j,\varsigma_l^*)=\eta_{lj}.
\end{equation}
By explicitly computing Eqs.~\eqref{R-rels} (and omitting the
product symbol $\cdot_F$), one obtains the relations stated in the
lemma. We denote by $\A[\C^4_\theta]$ the algebra generated by
$\{z_j,z_j^*~|~j=1,\ldots,4\}$ modulo the relations
\eqref{thetarels}. In this way, we have that $\A[\C^4_\theta]$ is a
left $H_F$-comodule $*$-algebra.\endproof

\begin{lem}
The algebra relations in the $H$-comodule algebra $\A[\RR^4]$ are
twisted into
\begin{equation}\label{th-st-rels}
\zeta_1\zeta_2 = \lambda \zeta_2\zeta_1,\quad
\zeta_1^*\zeta_2^*=\lambda\zeta_2^*\zeta_1^*, \quad
\zeta_2^*\zeta_1=\lambda\zeta_1\zeta_2^*,\quad
\zeta_2\zeta_1^*=\lambda\zeta_1^*\zeta_2,
\end{equation}
where the deformation parameter is $\lambda:=\mu^2=e^{2\pi\ii\theta}$.
\end{lem}

\proof The product on $\A[\RR^4]$ is once again twisted using the
formula \eqref{com-prod}. Again omitting the product symbol
$\cdot_F$, the relations are computed to be as stated. We denote by
$\A[\RR^4_\theta]$ the algebra generated by $\zeta_1$, $\zeta_2$ and
their conjugates, subject to these relations. They make
$\A[\RR^4_\theta]$ into a left $H_F$-comodule $*$-algebra.\endproof

\begin{rem}\textup{Since the generators $z_1$, $z_2$ and their
conjugates generate a commutative subalgebra of $\A[\C^4_\theta]$,
it is easy to see using Lemma~\ref{le:triv} that it is only the base
space $\RR^4_\theta$ of the localised twistor bundle that is
deformed. The typical fibre $\CP^1$ remains classical and the
localised twistor algebra is isomorphic to the tensor product
$\A[\RR^4_\theta]\otimes\A[\CP^1]$.}\end{rem}

The canonical differential calculi described in \S\ref{se:diffcal}
are also deformed. The relations in the quantised calculi are given
in the following lemmata.

\begin{lem}
The twisted differential calculus $\Omega(\C^4_\theta)$ is generated
by the degree zero elements $z_j, z_l^*$ and the degree one elements
$\D z_j, \D z_l^*$ for $j,l=1,\ldots,4$, subject to the bimodule
relations between functions and one-forms
\begin{align*}
z_j\D z_l=\eta_{lj}(\D z_l)z_j, & \qquad z_j\D z_l^*=\eta_{jl}(\D
z_l^*)z_j
\end{align*}
for $j,l=1,\ldots,4$ and the anti-commutation relations between
one-forms
\begin{align*}
\D z_j \wedge \D z_l + \eta_{lj}\D z_l \wedge \D z_j=0, & \qquad \D
z_j \wedge \D z_l^*+\eta_{jl}\D z_l^* \wedge \D z_j=0
\end{align*}
for $j,l=1,\ldots,4$.
\end{lem}

\proof One views the classical calculus $\Omega(\C^4)$ as a left
$H$-comodule algebra and accordingly computes the deformed product
using the twisting cocycle $F$. Since the exterior derivative $\D$
is $H$-equivariant, the (anti-) commutation relations in the
deformed calculus $\Omega(\C^4_\theta)$ are exactly the same as the
algebra relations in $\A[\C^4_\theta]$ but with $\D$ inserted
appropriately.\endproof

\begin{lem}\label{le:cl-calc}
The twisted differential calculus $\Omega(\RR^4_\theta)$ is
generated by the degree zero elements
$\zeta_1,\zeta_1^*,\zeta_2,\zeta_2^*$ and the degree one elements
$\D\zeta_1,\D\zeta_1^*,\D\zeta_2,\D\zeta_2^*$, subject to the
relations
\begin{align*}
\zeta_1 \, \D \zeta_2-\lambda\D \zeta_2 \, \zeta_1&=0, & \zeta_2^* \D \zeta_1 -\lambda \D \zeta_1 \, \zeta_2^*&=0, \\
\D \zeta_1 \wedge \D \zeta_2 + \lambda\D\zeta_2\wedge \D \zeta_1 &=
0, & \D \zeta_2^* \wedge \D \zeta_1 + \lambda\D\zeta_1\wedge \D
\zeta_2^* &= 0.
\end{align*}
\end{lem}

\proof Once again, the classical calculus $\Omega(\RR^4)$ is
deformed as a twisted left $H$-comodule algebra, with the relations
working out to be as stated.\endproof

In particular, it is clear that the vector space
$\Omega^2(\RR^4_\theta)$ is the same as it is classically. The Hodge
operator $*:\Omega^2(\RR^4)\to\Omega^2(\RR^4)$ commutes with the
$H$-coaction in the sense that
$$
\Delta_\pi(*\omega)=(\id\otimes *)\Delta_\pi(\omega),\qquad
\omega\in\Omega^2(\RR^4),
$$
so that there is also a Hodge operator
$*_\theta:\Omega^2(\RR^4_\theta)\to\Omega^2(\RR^4_\theta)$ defined
by the same formula as it is classically. There is a decomposition
of $\Omega^2(\RR^4_\theta)$ into self-dual and anti-self-dual
two-forms
$$
\Omega^2(\RR^4_\theta)=\Omega^2_+(\RR^4_\theta)\oplus\Omega^2_-(\RR^4_\theta)
$$
which, at the level of vector spaces, is identical to the
corresponding decomposition in the classical case.

We also apply the cocycle deformation the coordinate algebra
$\A[{\sf M}_k]$ of the space of self-conjugate monads by viewing it
as a left $H$-comodule algebra. We write $\A[{\sf M}_{k;\theta}]$
for the resulting cocycle-twisted left $H_F$-comodule algebra.

\begin{prop}\label{pr:cl-mon}
The noncommutative $*$-algebra $\A[{\sf M}_{k;\theta}]$ is generated
by the matrix elements $M^j_{ab}$, $N^l_{dc}$ for $a,c=1,\ldots,k$
and $b,d=1,\ldots,2k+2$, modulo the relations
$$
M^j_{ab}M^l_{cd}=\eta_{lj}M^l_{cd}M^j_{ab},\qquad
N^j_{ba}N^l_{dc}=\eta_{lj}N^l_{dc}N^j_{ba},
$$
together with the $*$-structure \eqref{monad star}.
\end{prop}

\proof From Lemma~\ref{le:mor} we read off the $H$-coaction on
generators $M^j$, $j=1,\ldots,4$, obtaining
\begin{align*}
M^j_{ab}&\mapsto \varsigma_j^*\otimes M^j_{ab}, \qquad
N^l_{dc}\mapsto \varsigma_l^*\otimes N^l_{dc},
\end{align*}
which we extend as a $*$-algebra map. The deformed relations follow
immediately from an application of the twisting formula
\eqref{com-prod}. The coaction of $H$ on $\A[{\sf M}_{k}]$ does not
depend on the matrix indices of the generators $M^j$, $N^l$,
$j,l=1,\ldots,4$, hence neither do the twisted commutation
relations. In terms of the deformed product, the relations
\eqref{adhm eqs} are twisted into the relations
$$
\sum\nolimits_r \left(N^j_{dr}M^l_{r
b}+\eta_{jl}N^l_{dr}M^j_{rb}\right)=0
$$
for all $j,l=1,\ldots,4$ and $b,d=1,\ldots,k$.\endproof

\subsection{The construction of instantons on $\RR^4_\theta$}
Just as we did for the Moyal plane, we now use the noncommutative
space of monads $\M_{k;\theta}$ to construct families of instantons on the
Connes-Landi space-time $\RR^4_\theta$.

The Pontryagin dual of the torus $\TT^2$ is the discrete group
$\widehat\TT^2\simeq \ZZ^2$. Given a pair of integers
$(r_1,r_2)\in\ZZ^2$ we define unitary elements $\vec
u=(u_1,u_2,u_3,u_4)$ of the algebra $H_F$ by
\begin{equation}\label{units2}
\vec
u=(u_1,u_2,u_3,u_4)=(\varsigma_1^{m_1},\varsigma_2^{m_2},\varsigma_3^{m_3},\varsigma_4^{m_4}),
\end{equation}
where $(m_j)=(r_1,r_1,r_2,r_2)$. It is clear that $u_1^*=u_2$ and
$u_3^*=u_4$, and that each $u_j$ is a group-like element of the Hopf
algebra $H_F$, {\em i.e.} it transforms as $\Delta(u_j)=u_j\otimes
u_j$ under the coproduct $\Delta:H_F\to H_F\otimes H_F$.

\begin{lem}\label{le:th-can}
There is a canonical left action of $H_F$ on the algebra
$\A[{\sfM}_{k;\theta}]$ defined on generators by
\begin{align*}
u_l\tr
M^j_{ab}&=\mathcal{R}(\varsigma_j^*,\varsigma_l^{m_l})M^j_{ab}=\eta_{lj}^{m_l}M^j_{ab},
& u_l\tr
M^j_{ab}{}^*&=\mathcal{R}(\varsigma_j,\varsigma_l^{m_l})M^j_{ab}{}^*=\eta_{jl}^{m_l}M^j_{ab}{}^*
\end{align*}
for $j,l=1,\ldots,4$.
\end{lem}

\proof From Proposition~\ref{pr:moyal-mon} we know that
$\A[{\sfM}_{k;\theta}]$ is a left $H_F$-comodule algebra; it is
therefore also a left $H_F$-module algebra according to the formula
\eqref{leftact}, which works out to be as stated.\endproof

This also gives us an action of the group $\ZZ^2$ on the algebra
$\A[{\sfM}_{k;\theta}]$ by $*$-automorphisms,
\begin{equation}\label{th-cross}
\gamma:\ZZ^2\to \mathrm{Aut}\,\A[{\sfM}_{k;\theta}].
\end{equation}
The smash product algebra $\A[{\sfM}_{k;\theta}]\lcross H_F$ corresponding the to $H_F$-action of Lemma~\ref{le:th-can} works out using the coproduct
$\Delta(\varsigma_j)=\varsigma_j\otimes \varsigma_j$ on $H_F$ and
the formula \eqref{cross} to have relations of the form
\begin{align*}
(M^j_{ab}\otimes u_l)(M^r_{cd}\otimes
u_s)&=\eta_{lr}^{m_l}\eta_{rj}\eta_{js}^{m_s}(M^r_{cd}\otimes
u_s)(M^j_{ab}\otimes u_l)\\ (M^j_{ab}\otimes
u_l)(M^r_{cd}{}^*\otimes
u_s)&=\eta_{rl}^{m_l}\eta_{rj}\eta_{sj}^{m_s}(M^r_{cd}{}^*\otimes
u_s)(M^j_{ab}\otimes u_l)
\end{align*}
for $j,l,r,s=1,\ldots,4$, together with their conjugates. This is
another special case of Example~\ref{ex:smash}. We think of this smash product
$\A[{\sfM}_{k;\theta}]\lcross H_F$ as an algebraic version the
crossed product algebra $\A[{\sfM}_{k;\theta}]\lcross_\gamma\,\ZZ^2$.

\begin{lem}
The algebra structure of the tensor product
$\A[{\sfM}_{k;\theta}]\utimes\A[\C^4_\theta]$ is determined by the
relations in the respective subalgebras $\A[{\sfM}_{k;\theta}]$ and
$\A[\C^4_\theta]$ given above, together with the cross-relations
\begin{align*}
M^jz_l=\eta_{jl}z_lM^j,\qquad M^jz_l^*=\eta_{lj}z_l^*M^j, \qquad
j,l=1,\ldots,4,
\end{align*}
as well as their conjugates.
\end{lem}

\proof  The classical algebra $\A[{\sf M}_k]\otimes\A[\C^4]$ is
equipped with the tensor product $H_F$-coaction of Eq.~\eqref{mon}.
We deform the product in this algebra using the formula
\eqref{com-prod}. The cross-terms in the resulting algebra
$\A[{\sfM}_{k;\theta}]\utimes\A[\C^4_\theta]$ are computed to be as
stated \cite{bl:mod}. Once again, the  symbol $\utimes$ is to remind
us that the algebra structure on the tensor product is not the
standard one and has been twisted by the deformation
procedure.\endproof

Once again we have a pair of matrices $\sigma_z$ and $\tau_z$,
\begin{align*}
\sigma_z&=\sum_j M^j\otimes z_j, & \tau_z&=\sum_j N^j\otimes z_j,
\end{align*}
but whose entries live in the twisted algebra
$\A[{\sfM}_{k;\theta}]\utimes\A[\C^4_\theta]$. The matrix
$\sfV:=\begin{pmatrix}\sigma_z&\sigma_{J(z)}\end{pmatrix}$ is a
$2k\times(2k+2)$ matrix with entries in
$\A[{\sfM}_{k;\theta}]\utimes\A[\C^4_\theta]$, using which we define
$\rho^2:=\sfV^*\sfV$. From the projection $\Qp:=\sfV\rho^{-2}\sfV^*$
we construct the complementary matrix $\Pp:=\mathbbm{1}_{2k+2}-\Qp$,
which has entries in the algebra
$\A[{\sfM}_{k;\theta}]\utimes\A[\RR^4_\theta]$.

It is clear that this matrix $\Pp$ is a self-adjoint idempotent,
$\Pp^2=\Pp=\Pp^*$. However, just as was the case for the Moyal
plane, it does not define an honest family of projections in the
sense of Definition~\ref{de:fam-bun}, since it has values in the
{\em twisted} tensor product algebra. We recover a
genuine family of projections using the following lemma, in which we
use the Sweedler notation $Z\mapsto Z\mo\otimes Z\bz$ for the left
coaction $\A[\C^4_\theta]\to H_F\otimes \A[\C^4_\theta]$ defined in
Eq.~\eqref{H-coact2}.

\begin{lem}
There is a canonical $*$-algebra map
$$
\mu:\A[{\sfM}_{k;\theta}]\utimes\A[\C^4_\theta]\to\left(\A[{\sfM}_{k;\theta}]\lcross
H_F\right)\otimes\A[\C^4_\theta]
$$
defined by $\mu(M\otimes Z)=M\otimes Z\mo\otimes Z\bz$ for each
$M\in \A[{\sfM}_{k;\theta}]$ and $Z\in \A[\C^4_\theta]$.
\end{lem}

\proof The proof is identical to that of Lemma~\ref{le:beta}, save
for the replacement of the coaction\eqref{tr-co1} by the coaction
\eqref{H-coact2}.\endproof

As a consequence, we find that there are maps
\begin{align}\label{tildesigma2}
\tilde\sigma_z:\mH\otimes\A[\C^4_\theta]&\to
\left(\A[{\sfM}_{k;\theta}]\lcross H_F\right)\otimes
\mK\otimes\A[\C^4_\theta],\\
\label{tildetau2}\tilde\tau_z:\mK\otimes\A[\C^4_\theta]&\to
\left(\A[{\sfM}_{k;\theta}]\lcross H_F\right)\otimes
\mL\otimes\A[\C^4_\theta]
\end{align}
defined by composing $\sigma_z$ and $\tau_z$ with the map $\mu$.
With the coaction \eqref{H-coact2}, they work out to be
\begin{align*}
\tilde\sigma_z&:=\sum_r M^r\otimes \varsigma_r\otimes z_r, &
\tilde\tau_z&:=\sum_r N^r\otimes \varsigma_r\otimes z_r,
\end{align*}
which are respectively $k\times (2k+2)$ and $(2k+2)\times k$
matrices with entries in the noncommutative algebra
$(\A[{\sfM}_{k;\theta}]\lcross H_F)\otimes\A[\C^4_\theta]$. With
this in mind, we form the $(2k+2)\times 2k$ matrix
$\tsfV:=\begin{pmatrix}\tilde\sigma_z&\tilde\sigma_{J(z)}\end{pmatrix}$,
this time yielding a $2k\times(2k+2)$ matrix with entries in
$(\A[{\sfM}_{k;\theta}]\lcross H_F)\otimes\A[\C^4_\theta]$, and
define $\trho^2:=\tsfV^*\tsfV$. Just as in the classical case, in
order to proceed we need to slightly enlarge the matrix algebra
$\M_k(\C)\otimes\A[{\sf M}_{k;\theta}]\otimes \A[\C^4_\theta]$ by
adjoining an inverse element $\trho^{-2}$ for $\trho^2$.

\begin{prop}
The $(2k+2)\times(2k+2)$ matrix $\tQp=\tsfV\trho^{-2}\tsfV^*$ is a
projection, $\tQp^2=\tQp=\tQp^*$, with entries in the algebra
$(\A[{\sfM}_{k;\theta}]\lcross H_F)\otimes\A[\C^4_\theta]$ and trace
equal to $2k$.
\end{prop}

\proof The fact that $\tQp$ is a projection follows from the fact
that $\Qp$ is a projection and $\mu$ is a $*$-algebra map. By
construction, the entries of the matrix $\trho^2$ are central in the
algebra $(\A[{\sfM}_{k;\theta}]\lcross H_F)\otimes\A[\C^4_\theta]$
(this follows from the fact that the corresponding classical matrix
elements are coinvariant under the left $H$-coaction), from which it
follows that the trace computation in Proposition~\ref{pr:adhm-pr}
is valid in the noncommutative case as well \cite{bl:mod}.\endproof

From the projection $\tQp$ we construct the complementary projection
$\tPp:=\mathbbm{1}_{2k+2}-\tQp$; it has entries in the algebra
$(\A[{\sfM}_{k;\theta}]\lcross H_F)\otimes\A[\RR^4_\theta]$ and has
trace equal to two. In analogy with Definition~\ref{de:fam-bun}, the
finitely-generated projective module
$$
\E:=\tPp\left((\A[{\sfM}_{k;\theta}]\lcross
H_F)\otimes\A[\RR^4_\theta]\right)^{2k+2}
$$
defines a family of rank two vector bundles over $\RR^4_\theta$
parameterised by the noncommutative algebra
$\A[{\sfM}_{k;\theta}]\lcross H_F$. We equip this family of vector
bundles with the family of Grassmann connections associated to the
projection $\tPp$.

\begin{prop}
The curvature $F=\tPp((\id\otimes \D)\tPp)^2$ of the Grassmann
family of connections $\n:=(\id\otimes \D)\circ\tPp$ is
anti-self-dual.
\end{prop}

\proof From Lemma~\ref{le:stcalc} we know that the space of
two-forms $\Omega^2(\RR^4_\theta)$ and the Hodge $*$-operator
$*_\theta:\Omega^2(\RR^4_\theta)\to\Omega^2(\RR^4_\theta)$ are
undeformed and equal to their classical counterparts; similarly for
the decomposition
$\Omega^2(\RR^4_\theta)=\Omega^2_+(\RR^4_\theta)\oplus\Omega^2_-(\RR^4_\theta)$
into self-dual and anti-self-dual two-forms. This identification of
the `quantum' with the `classical' spaces of two-forms survives the
tensoring with the parameter space $\A[{\sfM}_{k;\theta}]\lcross
H_F$, which yields that $(\A[{\sfM}_{k;\theta}]\lcross
H_F)\otimes\Omega^2_\pm(\RR^4_\theta)$ and $(\A[{\sfM}_{k}]\otimes
H)\otimes\Omega^2_\pm(\RR^4)$ are isomorphic as vector spaces.
Computing the curvature $F$ in exactly the same way as in
Proposition~\ref{pr:asd}, we see that it must be anti-self-dual,
since the same is true in the classical case.\endproof

\subsection{The toric ADHM equations}
The previous section produced a family of instantons parameterised
by the noncommutative algebra $\A[{\sfM}_{k;\theta}]\lcross H_F$.
Just as we did for the Moyal space-time, we would like to find a
suitable commutative subalgebra of $\A[{\sfM}_{k;\theta}]\lcross
H_F$ and hence a family of instantons parameterised by a classical
space. In order to do this, we introduce elements of
$\A[{\sfM}_{k;\theta}]\lcross H_F$ defined by
\begin{align*}
\widetilde M^1_{ab}&:=M^1_{ab}\otimes \varsigma_1, & \widetilde
M^2_{ab}&:=M^2\otimes \varsigma_2, & \widetilde
M^3_{ab}&:=M^3_{ab}\otimes 1, & \widetilde
M^4_{ab}&:=M^4_{ab}\otimes 1
\end{align*}
for each $a=1,\ldots,k$ and $b=1,\ldots,2k+2$, together with their
conjugates $\widetilde M^j_{ab}{}^*$, $j=1,\ldots,4$.

\begin{defn}
We write $\A[\mathfrak{M}(k;\theta)]$ for the $*$-subalgebra of
$\A[{\sfM}_{k;\theta}]\lcross H_F$ generated by the elements
$\widetilde M^j_{ab}{}$, $\widetilde M^l_{ab}{}^*$.
\end{defn}

\begin{prop}
The algebra $\A[\mathfrak{M}(k;\theta)]$ is a commutative
$*$-subalgebra of the smash product $\A[{\sfM}_{k;\theta}]\lcross
H_F$.
\end{prop}

\proof The generators $\widetilde M^3_{ab}$, $\widetilde M^4_{ab}$
and their conjugates are obviously central. The generators
$\widetilde M^1_{ab}$, $\widetilde M^2_{ab}$ and their conjugates
are also easily seen to commute amongst themselves. We check the
case $j\in\{1,2\}$, $l\in \{3,4\}$, yielding
\begin{align*}
\widetilde M^j_{ab}\widetilde
M^l_{cd}&=(M^j_{ab}\otimes\varsigma_j)(M^l_{ab}\otimes
1)=\eta_{jl}M^j_{ab}M^l_{cd}\otimes\varsigma_j=\eta_{jl}\eta_{lj}M^l_{cd}M^j_{ab}\otimes\varsigma_j\\
&=(M^l_{cd}\otimes 1)(M^j_{ab}\otimes\varsigma_j)=\widetilde
M^l_{cd}\widetilde M^j_{ab}.
\end{align*}
All other pairs of generators are shown to commute using similar
computations.\endproof

Although we have changed our set of generators, we nevertheless combine
them with the Hopf algebra $H_F$ using a smash product construction.
Let $\tr':H_F\otimes\A[\mathfrak{M}(k;\theta)]\to
\A[\mathfrak{M}(k;\theta)]$ be the left $H_F$-action defined by
\begin{align*}
\varsigma_l\tr' \widetilde M^j_{ab}&=\eta_{lj}\widetilde M^j_{ab}, &
\varsigma_l\tr' \widetilde M^j_{ab}{}^*&=\eta_{jl}\widetilde
M^j_{ab}{}^*
\end{align*}
for $j,l=1,\ldots,4$ and let $\A[\mathfrak{M}(k;\theta)]\lcross H_F$
be the corresponding smash product algebra. The next proposition
relates the parameter space $\A[\mathfrak{M}(k;\theta)]$ to the
parameter space $\A[{\sfM}_{k;\theta}]$.

\begin{thm}
There is a $*$-algebra isomorphism
$\phi:\A[\mathfrak{M}(k;\theta)]\lcross H_F\to
\A[{\sfM}_{k;\theta}]\lcross H_F$ defined for each $h\in H_F$ by
\begin{align*}
\widetilde M^1_{ab}\otimes h&\mapsto M^1_{ab}\otimes\varsigma_1h, &
\widetilde M^2_{ab}\otimes h&\mapsto M^2_{ab}\otimes\varsigma_2h, \\
\widetilde M^3_{ab}\otimes h&\mapsto M^3_{ab}\otimes h, & \widetilde
M^4_{ab}\otimes h&\mapsto M^4_{ab}\otimes h
\end{align*}
and extended as a $*$-algebra map.
\end{thm}

\proof The given map is clearly a vector space isomorphism with
inverse
\begin{align*}
M^1_{ab}\otimes h&\mapsto \widetilde M^1_{ab}\otimes\varsigma_1^*h,
& M^2_{ab}\otimes h&\mapsto \widetilde
M^2_{ab}\otimes\varsigma_2^*h, \\ M^3_{ab}\otimes h&\mapsto
\widetilde M^3_{ab}\otimes h, & M^4_{ab}\otimes h&\mapsto \widetilde
M^4_{ab}\otimes h,
\end{align*}
extended as a $*$-algebra map. By definition, the map $\phi$ is a
$*$-algebra homomorphism on the subalgebra
$\A[\mathfrak{M}(k;\theta)]$, so it remains to check that it
preserves the cross-relations with the subalgebra $H_F$. This is
easy to verify: one has for example that
\begin{align*}
\phi(1\otimes \varsigma_j)\phi(\widetilde M^1\otimes 1)&=(1\otimes
\varsigma_j)(M^1\otimes \varsigma_1)
=\eta_{j1}M^1\otimes \varsigma_j\varsigma_1\\
&=\eta_{j1}(M^1\otimes\varsigma_1)(1\otimes \varsigma_j)
=\eta_{j1}\phi(\widetilde M^1\otimes 1)\phi(1\otimes\varsigma_j).
\end{align*}
The remaining relations are checked in exactly the same
way.\endproof

Next we focus on the task of seeing how the parameters corresponding
to the subalgebra $H_F$ can be removed in order to leave a family of
instantons parameterised by the commutative algebra
$\A[\mathfrak{M}(k;\theta)]$. There is a right coaction
\begin{equation}\label{gauge-co2}
\delta_R:\A[\mathfrak{M}(k;\theta)]\lcross H_F\to
\left(A[\mathfrak{M}(k;\theta)]\lcross H_F\right)\otimes H_F,\qquad
\delta_R:=\id\otimes\Delta,
\end{equation}
where $\Delta:H_F\to H_F\otimes H_F$ is the coproduct on the Hopf
algebra $H_F$. This coaction is by gauge transformations, meaning
that that the projections $\widetilde\Pp\otimes 1$ and
$\delta_R(\widetilde\Pp)$ are unitarily equivalent in the matrix
algebra $\M_{2k+2}\left((\A[\mathfrak{M}(k;\theta)]\lcross
H_F)\otimes H_F\right)$ and so they define gauge equivalent families
of instantons \cite{bl:mod}. 

The parameters determined by the
subalgebra $H_F$ in $\A[\mathfrak{M}(k;\theta)]\lcross H_F$ are
therefore just `gauge' parameters and so they may be removed by
passing to the subalgebra of $\A[\mathfrak{M}(k;\theta)]\lcross H_F$
consisting of coinvariant elements under the coaction
\eqref{gauge-co2}, {\em viz}.
$$
\A[\mathfrak{M}(k;\theta)]\cong\{a\in\A[\mathfrak{M}(k;\theta)]\lcross
H_F~|~\delta_R(a)=a\otimes 1\}.
$$
In this way we obtain a projection $\Pp_{k;\theta}$ with entries in
$\A[\mathfrak{M}(k;\theta)]\otimes\A[\RR^4_\theta]$. The explicit
details of the construction of the projection $\Pp_{k;\theta}$ are
given in \cite{bl:mod}, together with a proof of the fact that the
Grassmann family of connections $\n=\Pp_{k;\theta}\circ(\id\otimes
\D)$ also has anti-self-dual curvature and hence defines a family of
instantons on $\RR^4_\theta$.

Moreover, the commutative algebra $\A[\mathfrak{M}(k;\theta)]$ is
the algebra of coordinate functions on a classical space of monads
$\mathfrak{M}(k;\theta)$. For each point
$x\in\mathfrak{M}(k;\theta)$ there is an evaluation map
$$
\textrm{ev}_x\otimes\id:\A[\mathfrak{M}(k;\theta)]\otimes\A[\C^4_\theta]\to
\A[\C^4_\theta],
$$
which in turn determines monad over the noncommutative space
$\C^4_\theta$ in terms of the matrices $(\textrm{ev}_x\otimes\id)\widetilde
\sigma_z$ and $(\textrm{ev}_x\otimes\id)\widetilde \tau_z$, {\em i.e.} a
sequence of free right $\A[\C^4_\theta]$-modules
\begin{equation}\label{monad-theta}
0\to \mH\otimes
\A[\C^4_\theta]\xrightarrow{(\textup{ev}_x\otimes\id)\tilde\sigma_z}
\mK\otimes \A[\C^4_\theta]
\xrightarrow{(\textup{ev}_x\otimes\id)\tilde\tau_z} \mL\otimes
\A[\C^4_\theta]\to 0.
\end{equation}
Recall from Proposition~\ref{pr:gauge} that the gauge freedom in the
classical ADHM construction is precisely the freedom to choose
linear bases of the the vector spaces $\mH$, $\mK$, $\mL$. Clearly
we also have this freedom in the noncommutative construction and so we write $\sim$ for the resulting equivalence relation on the space $\mathfrak{M}(k;\theta)$ ({\em cf}. Definition~\ref{de:mon-eq}). This yields the following description of the space
$\mathfrak{M}(k;\theta)$ of classical parameters in the ADHM construction on $\RR^4_\theta$.

\begin{thm}\label{th:tor-adhm}
For $k\in \ZZ$ a positive integer, the space
$\mathfrak{M}(k;\theta)/\sim$ of equivalence classes of self-conjugate monads over $\C^4_\theta$
is the quotient of the set of complex matrices $B_1,B_2\in\M_k(\C)$,
$I\in \M_{2\times k}(\C)$, $J\in\M_{k\times 2}(\C)$ satisfying the
equations
\begin{enumerate}[\hspace{0.5cm}(i)]
\item $\bar\mu B_1B_2-\mu B_2B_1+IJ=0$,
\item
$[B_1,B_1^*]+[B_2,B_2^*]+II^*-J^*J=0$
\end{enumerate}
by the action of $\U(k)$ given by
\begin{equation*}
B_1\mapsto gB_1g^{-1},\qquad B_2\mapsto gB_2g^{-1}, \qquad I\mapsto
gI,\qquad J\mapsto Jg^{-1}
\end{equation*}
for each $g\in\U(k)$.
\end{thm}

\proof We express the monad maps $\tilde\sigma_z$, $\tilde\tau_z$ as
$$
\tilde\sigma_z=\widetilde M^1z_1+\widetilde M^2z_2+\widetilde
M^3z_3+\widetilde M^4z_4,\qquad \tilde\tau_z=\widetilde
N^1z_1+\widetilde N^2z_2+\widetilde N^3z_3+\widetilde N^4z_4
$$
for constant matrices $\widetilde M^j$, $\widetilde N^l$,
$j,l=1,\ldots,4$. Upon expanding out the condition
$\tilde\tau_z\circ\tilde\sigma_z=0$ and using the commutation
relations in Lemma~\ref{le:sttheta}, we find the conditions
\begin{equation}\label{tor-mc}
\widetilde N^j\widetilde M^l+\eta_{jl}\widetilde N^l\widetilde M^j=0
\end{equation}
for $j,l=1,\ldots,4$. The typical fibre $\CP^1$ of the twistor
fibration $\RR^4\times\CP^1$ has homogeneous coordinates $z_1$,
$z_1^*$, $z_2$, $z_2^*$ and the `line at infinity' $\ell_\infty$ is
recovered by setting $z_1=z_2=0$. On this line, the monad condition
$\tilde\tau_z\circ\tilde\sigma_z=0$ becomes
\begin{equation}\label{inf}
\widetilde N^3\widetilde M^4+\widetilde N^4\widetilde M^3=0, \qquad
\widetilde N^3\widetilde M^3=0,\qquad \widetilde N^4\widetilde
M^4=0.
\end{equation}
Moreover, when $z_1=z_2=0$ we see from the relations
\eqref{thetarels} that the coordinates $z_3$, $z_4$ and their
conjugates are mutually commuting, so that the line $\ell_\infty$ is
classical. Self-conjugacy of the monad once again implies that the
restricted bundle over $\ell_\infty$ is trivial, whence we can argue
as in \cite{oss:vb} to show that the map $\widetilde N^3\widetilde
M^4=-\widetilde N^4\widetilde M^3$ is an isomorphism. We choose
bases for $\mH,\mK,\mL$ such that $\widetilde N^3\widetilde
M^4=\mathbbm{1}_k$ and
\begin{align*}
\widetilde M^3&=\begin{pmatrix}\mathbbm{1}_{k\times k} \\ 0_{k\times k} \\
0_{2\times k}\end{pmatrix}, & \widetilde M^4&=\begin{pmatrix}0_{k\times k} \\\mathbbm{1}_{k\times k}\\
0_{2\times k}\end{pmatrix}, & \widetilde
N^3&=\begin{pmatrix}0_{k\times k} \\ \mathbbm{1}_{k\times k} \\
0_{k\times 2}\end{pmatrix}^{\textrm{tr}}, & \widetilde
N^4&=\begin{pmatrix}-\mathbbm{1}_{k\times k} \\ 0_{k\times k} \\
0_{k\times 2}\end{pmatrix}^{\textrm{tr}}.
\end{align*}
Using the conditions \eqref{tor-mc} for $j=3,4$ and $l=1,2$, the
remaining matrices are necessarily of the form
\begin{align*}
\widetilde M^1&=\begin{pmatrix}B_1\\B_2\\J\end{pmatrix}, &
\widetilde M^2&=\begin{pmatrix}B_1'\\B_2'\\J'\end{pmatrix}, &
\widetilde N^1&=\begin{pmatrix}-\mu B_2 \\ \bar\mu B_1
\\ I\end{pmatrix}^{\textrm{tr}}, & \widetilde N^2&=\begin{pmatrix}-\bar\mu B_2' \\ \mu
B_1' \\ I' \end{pmatrix}^{\textrm{tr}}.
\end{align*}
Invoking the relations $\tilde\tau_{J(z)}^*=-\tilde\sigma_z$ and
$\tilde\sigma_{J(z)}^*=\tilde\tau_z$ corresponding to the fact that
the monad is self-conjugate, we find that
$$
B_1'=-\bar\mu B_2^*,\qquad B_2'=\mu B_1^*,\qquad J'=I^*,\qquad
I'=-J^*.
$$
Thus in order to fulfil the condition
$\tilde\tau_z\circ\tilde\sigma_z=0$ it remains to impose conditions
\eqref{tor-mc} in the cases $j=l=1$ and $j=1,l=2$. These are
precisely conditions (i) and (ii) in the theorem. It is evident just
as in the classical case \cite{don} that the remaining freedom in this set-up
is given by the stated action of $\U(k)$, whence the result.
\endproof

\medskip
\begin{center}
\textsc{Acknowledgments}
\end{center}

SJB gratefully acknowledges support from the ESF network `Quantum
Geometry and Quantum Gravity' and the NWO grant 040.11.163. WvS acknowledges support from NWO under VENI-project 639.031.827. Both authors thank the Institut des Hautes \'Etudes Scientifiques for hospitality during a short visit in 2010.

\end{document}